\documentclass[12pt,preprint]{aastex6} 
\usepackage{subfigure}
\usepackage{graphicx}
\usepackage{amsmath}

\slugcomment{Draft, revised 1, \today}

\shorttitle{New Suns in the Cosmos IV}
\shortauthors{D. B. de Freitas et al.}

\begin{document}

\title{New Suns in the Cosmos IV: the multifractal nature of stellar magnetic activity in \textit{Kepler} cool stars}

\author{D. B. de Freitas\altaffilmark{1}, M. M. F. Nepomuceno\altaffilmark{2,3}, M. Gomes de Souza\altaffilmark{2}, I. C. Le\~ao\altaffilmark{2,4}, M. L. Das Chagas\altaffilmark{5}, A. D. Costa\altaffilmark{2}, B. L. Canto Martins\altaffilmark{2} and J. R. De Medeiros\altaffilmark{2}}

\altaffiltext{1}{Departamento de F\'{\i}sica, Universidade Federal do Cear\'a, Caixa Postal 6030, Campus do Pici, 60455-900 Fortaleza, Cear\'a, Brazil}
\altaffiltext{2}{Departamento de F\'{\i}sica, Universidade Federal do Rio Grande do Norte, 59072-970 Natal, RN, Brazil}
\altaffiltext{3}{Universidade Federal Rural do Semi-\'Arido, Campus Central Costa e Silva, CEP 59600-900, Mossor\'o, RN, Brazil}
\altaffiltext{4}{European Southern Observatory, Karl-Schwarzschild-Str. 2, 85748 Garching, Germany}
\altaffiltext{5}{Faculdade de F\'{\i}sica - Instituto de Ci\^encias Exatas, Universidade Federal do Sul e Sudeste do Par\'a, Marab\'a, PA 68505-080, Brazil}

\begin{abstract}
In the present study, we investigate the multifractal nature of a long-cadence time series observed by the \textit{Kepler} mission for a sample of 34 M dwarf stars and the Sun in its active phase. Using the Multifractal Detrending Moving Average algorithm (MFDMA), which enables the detection of multifractality in nonstationary time series, we define a set of multifractal indices based on the multifractal spectrum profile as a measure of the level of stellar magnetic activity. This set of indices is given by the ($A$,$\Delta \alpha$,$C$,$H$)-quartet, where $A$, $\Delta \alpha$ and $C$ are related to geometric features from the multifractal spectrum and the global Hurst exponent $H$ describes the global structure and memorability of time series dynamics. As a test, we measure these indices and compare them with a magnetic index defined as $S_{ph}$ and verify the degree of correlation among them. First, we apply the Poincar\'e plot method and find a strong correlation between the $\left\langle S_{ph}\right\rangle$ index and one of the descriptors that emerges from this method. As a result, we find that this index is strongly correlated with long-term features of the signal. From the multifractal perspective, the $\left\langle S_{ph}\right\rangle$ index is also strongly linked to the geometric properties of the multifractal spectrum except for the $H$ index. Furthermore, our results emphasize that the rotation period of stars is scaled by the $H$ index, which is consistent with Skumanich's relationship. Finally, our approach suggests that the $H$ index may be related to the evolution of stellar angular momentum and a star's magnetic properties.
\end{abstract}

\keywords{stars: solar-type --- stars: astrophysical time series --- Sun: rotation --- methods: data analysis}

\section{Introduction}\label{sec:intro}
Stellar activity is strongly related to extreme and transient phenomena caused by the dissipation of magnetic energy when magnetic fields evolve from complex to simple topologies according to the rate of stellar rotation \citep{defreitasmnras,terry}. Indeed, as cited by \cite{saar}, the magnetic cycle period and rotation period are correlated in such a manner that slow rotators have larger magnetic cycles. Furthermore, the global level of magnetic activity changes over time following cycles, and the local levels change on different timescales that are associated with high and low intermittent fluctuations, which change on timescales ranging from a few seconds to several hours (flares) and from days to weeks (active regions), respectively. However, this level of activity depends on the rotation of stars. As mentioned by \cite{mathur2014b}, the most rapidly rotating stars show the shortest activity cycles, while more slowly rotating stars generally present cycle periods similar to the Sun or longer. Depending on the star mass, the variability in stellar behavior at small and large scales is a fundamental factor that can be used to better understand stellar magnetism phenomena \citep{defreitas2013,terry}. Because the magnetic activity that emerges from the stellar surface is a physical mechanism produced by the stellar dynamo, the long-term variations due to the rotational period play an important role in understanding the level of activity on the surface of the star. Specifically, for low-mass stars, the observed magnetism is associated with the external convective envelope, where strong mass motions of conductive plasma induce a magnetic field through a cyclic dynamo process \citep{brun2015}. On the other hand, fossil magnetism is not the result of these same mechanisms. This kind of magnetism arises from the trapping of the magnetic flux after stellar formation or an initial dynamo phase and may therefore be the mechanism for long-timescale stability without the dynamo effect \citep{duez,duez2010}. However, it is necessary to investigate the effects of magnetic activity as a source of noise responsible for stellar micro-variability when modeling different features in the frequency spectrum \citep{karoff}.

In particular, fully convective stars show granulation signatures on their surfaces \citep{karoff}. This type of target is particularly attractive because of different physical characteristics, particularly the transition from partially convective Sun-like stars to fully convective stars, likely around the spectral type M3/M4 \citep{reiners}. As \cite{Houdek} indicated, inside these stars, acoustic oscillations are excited by the outer convection zone, where their amplitudes at the surfaces can be observed. Thus, the signatures of both granulation and oscillations are included in the noise background of the time series at high frequencies. The signatures at low frequencies can be caused by the rotational modulation of long-lived sunspots \citep{lanza2004}.

For main-sequence stars, the study of stellar magnetic activity focuses on the analysis of signatures of fluctuations in short- and long-lived variations because of different physical processes, which manifest in characteristic timescales from oscillation ($>2000\mu$Hz) to rotation ($<1\mu$Hz) \citep{karoff}. These processes are generated by complex temporal dynamics in the stellar photosphere and atmosphere. In general, the mechanisms responsible for the magnetic activity, such as interactions among rotation, convection and the magnetic field, which generate the differential rotation, are not completely understood \citep{brun2011,daschagas2016}. In this case, it is necessary to investigate the statistical relationships among longer-term variations because of the stellar magnetism cycle and among shorter-term variations because of the correlated noises that differ from Gaussian noise \citep{cw2009}. It is important to highlight that although the short-term variations are dominated by granulation and oscillations, while the long-term variations are related to rotation and magnetism, the spectra of  M dwarfs are dominated by photon noise at approximately $200 \mu$Hz and above, depending on the magnitude of the star. For these targets, both the granulation noise and  oscillations are at much higher frequencies than the Nyquist limit. Hence, the power spectra of these targets at low frequencies are dominated by the rotation period and its harmonics and a power slope related to magnetic effects. As our proposal is to study M dwarfs using long-cadence \textit{Kepler} data, this brief review is very relevant and crucial in the context of this paper.

Since the pioneering results of \cite{wb}, a.k.a. the WB effect, several efforts have been made to measure the level of magnetic activity in photometric and spectroscopic observations. In general, this effect relates the absolute magnitude ($M_{V}$) and separation between the outer edges of the Ca\texttt{II} K emission line ($\log[W_{0}]$ in km/s) \citep[see \textit{e.g.},][]{terry}. From the empirical correlation $\log[W_{0}]-M_{V}$, a linear slope is extracted as a measure of the level of magnetic activity. Another magnetic activity proxy is the average Mount Wilson $S$ index proposed by \cite{Middelkoop}. The physical index $R_{HK}$ is also used to measure the magnetic index and is related to the $S$ index by $R_{HK} = C_{B-V} S$, where $C_{B-V}$ denotes a color-dependent correction factor \citep[cf.][]{Cincunegui}. \cite{noyes} investigated the chromospheric magnetic activity index $R_{HK}$ as a function of both rotation period and spectral type through the Rossby number $R_{o} = P_{rot}/\tau_{c}$, where $\tau_{c}$ is the convective turnover time and $P_{rot}$ denotes the rotational period. Recently, \cite{mathur2014a} suggested a new magnetic index based on the global $S_{ph}$ index, which \cite{garcia} defines as the standard deviation of the entire time series. Considering the magnetic activity fluctuations with time, \cite{mathur2014a} proposed the mean value of $\left\langle S_{ph}\right\rangle$ as a global magnetic activity index, which was divided into independent subseries of length $k\times P_{rot}$, where $k$ is an integer number set to 5 and $P_{rot}$ is the rotational period of the star. This procedure has the great advantage of accounting for the effects of rotation of the star, allowing for the measurement of the magnetic index during the minimum (or maximum) regime of the magnetic cycle. The next step taken by the authors was to correct the value of the $\left\langle S_{ph}\right\rangle$ index for photon noise. To this end, they made use of the magnitude correction from \cite{{jenkins2010}} to estimate the corrections to apply to the \textit{Kepler} stars. Hence, they ensure that $\left\langle S_{ph}\right\rangle$ is dominated by the magnetism of each star, eliminating any effects due to granulation. Furthermore, these authors assume this new index as a counterpart of the spectroscopic Mount Wilson $S$ index. 

Recently, \cite{defreitas2013} showed that the rotational periods of a sample of solar-type stars from the CoRoT database are linked to the Hurst index $H$ by a simple logarithmic relationship \citep[cf.][]{demedeiros2013}. More recently, \cite{defreitas2016} have shown that the level of complexity in stellar activity is associated with the degree of multifractality and the asymmetry of the multifractal spectrum. More specifically, the authors note that short time series are characterized by strong long-range correlations because of rotational modulation, where the flicker noise is used as a good candidate to explain the changes in the multifractal index. As mentioned by de Freitas et al. (2016), the flicker noise \citep{bastien} is the stellar granulation measured in the time domain that appears at lower frequencies and on timescales shorter than 8 hours. The granulation signal is widely used as a proxy for the stellar surface gravity \citep{mathur2011,bastien2016,kallinger2016}. Generally, this multifractal approach suggests that the growth of the rotation rate destroys the ``multifractal diversity'', which is given by the index $\Delta\alpha$ and indicates which different rotation regimes affect local structures \citep{aschwa,a2011}. In other words, this approach implies that a high long-term persistence (the longest period) has a low level of complexity; i.e., it has a strong system memorability \citep{tang}.

In Section 2, we first describe a geometric technique, a.k.a. Poincar\'e plot, that relates different levels of variability in short- to long-term periods. We also define a set of the magnetic activity indices from multifractal formalism based on the geometric features of the generalized fractal dimension spectrum. In this context, we use the statistical procedure suggested by \cite{defreitas2016}. In Section 3, we compare these indices with the average $S_{ph}$ index developed by \cite{mathur2014a} using a sample of 34 M notably low-mass dwarfs, which were observed by the \textit{Kepler} mission and initially treated by these authors. In this section, we also discuss the physical implications of the $S_{ph}$ index and its relationship with the geometric properties of the multifractal spectrum. In the final section, we present our final comments and conclusions.

\section{Statistical background}
Several methods are used for analyzing the behavior of the standard deviation (or variance) of a time series, among which we can point a self-similar statistical method known as the Poincar\'e plot \citep{Arkhypov}, as well as other methods that assume the time series as a self-affine fractal, known as Scaled Windowed Variance ($SWV$) analysis \citep{seuront} and Rescaled Range ($R/S$) analysis \citep{hurst1951,mw1969a,defreitas2013}. The degree of the robustness of the methods follows from the Poincar\'e plot to the $R/S$ analysis.

In this context, the method applied by \cite{mathur2014a} is, for purposes of fast inspection, more similar to the Poincar\'e plot. This conclusion is due to simple fact that the Poincar\'e plot analyzes only the global aspect of the fluctuations, while the $SWV$ and $R/S$ methods were developed for assessing the behavior of the fluctuations at different timescales. The intention of the authors was to find a metric adapted to each star to compare their magnetism instead of fixing the time scale and comparing their variability. In this way, the $S_{ph}$ index is a metric linked to spots and thus to the surface magnetism. It is noteworthy that the Poincar\'e plot is widely used in cases where the signal is stationary.

\subsection{Poincar\'e plot}

A time series $x(t)$ comprises a deterministic function $p(t,\bar{P})$, which represents a type of global trend, and a set of short-term variations, which are denoted by random walk  $r(t)$ and stationary noise $\eta(t)$ components; thus, $x(t)=p(t,\bar{P})+r(t)+\eta(t)$ \citep{defreitas2016}. A global trend can be modeled by nonlinear functions or smoothers, whereas a short-term structure can be treated using a differencing filter given by $x_{n+1}-x_{n}$, where $n$ denotes the number of points. According to \cite{feigelson}, the combination of these procedures constructs a return map, which simultaneously fits long-term deterministic trends and periodicities and short-term autocorrelated fluctuations. Owing to theoretical background from which self-similarity processes emerged, this method is named the \textit{Poincar\'e plot}.

\begin{figure*}
\begin{center}
\subfigure{\includegraphics[width=0.88\textwidth]{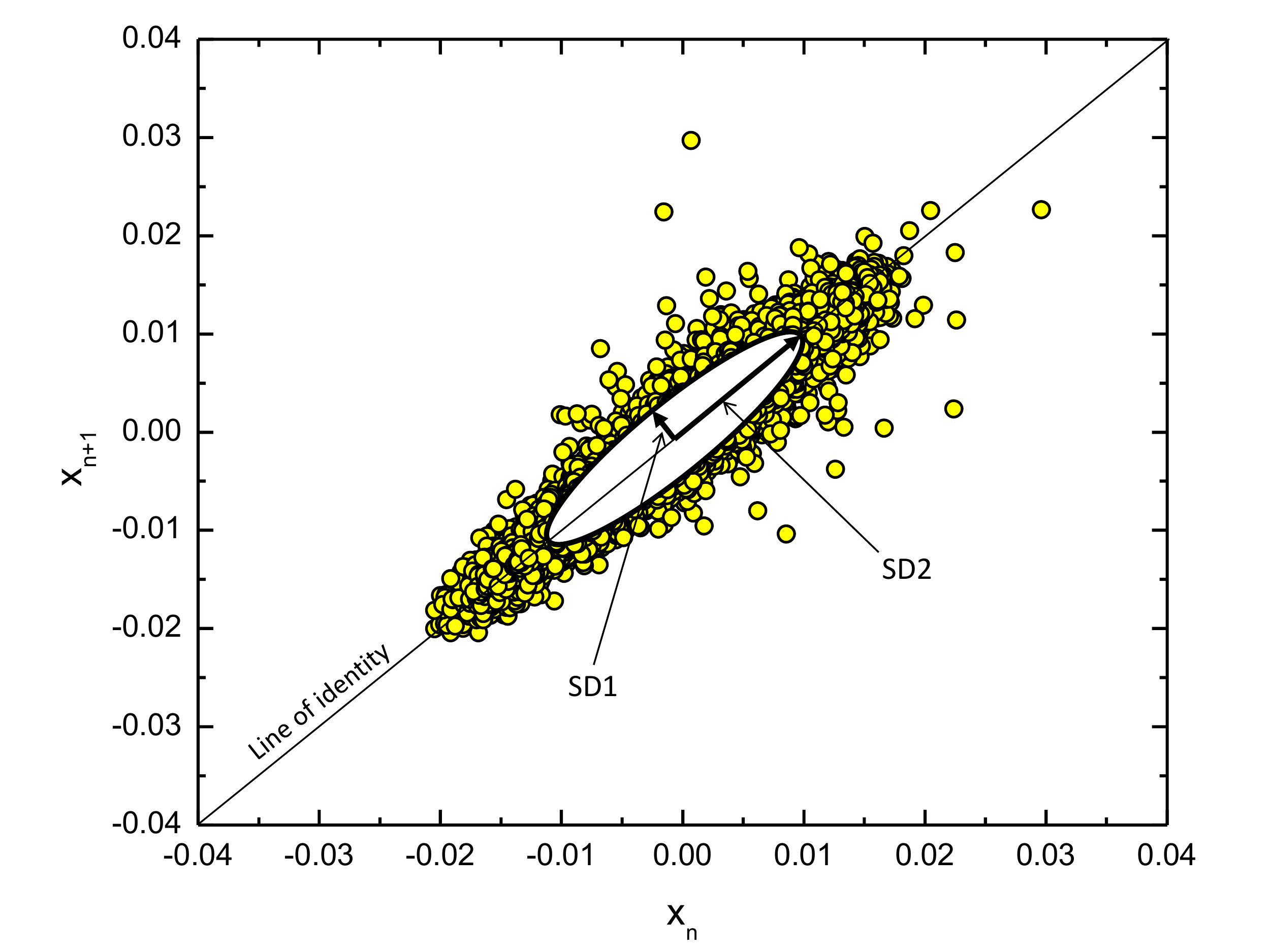}}
\end{center}
\caption{An illustration of an ellipse fitted to the Poincar\'e plot and descriptors SD1
and SD2.
}
\label{fig0a}
\end{figure*}

The Poincar\'e plot is a geometric method that analyzes the dynamic behavior of time series. This method represents a time series in a Cartesian plane, where each measurement result is plotted as a function of a previous result, and the result is similar to Fig. \ref{fig0a} \citep{tulppo}. Our study applies the procedure used by \cite{tulppo} to compute the Poincar\'e plot.

Strictly speaking, a Poincar\'e plot can be analyzed by adjusting an ellipse to the diagram formed by the attractor with its center at $(0,0)$, as shown in Fig.\ref{fig0a}. The SD1 line indicates the dispersion of data perpendicular to the identity line ($x_{n+1}=x_{n}$), whereas the standard deviation along the identity line is represented by SD2.

SD1 and SD2 standard descriptors are calculated by turning the Poincar\'e plot 45$^{o}$ clockwise and 45$^{o}$ counterclockwise, respectively. Roughly speaking, SD1 is the standard descriptor of the width histogram projected on the identity line, whereas SD2 is measured by the length histogram, which is obtained by projecting the points onto the identity line \textbf{\citep{brennan2001,brennan2002,Arkhypov}}. According to the traditional Poincar\'e analysis, SD1 is often used as a measure of the short-term variability, and SD2 is a measure of the long-term variability. However, these aspects are only correct when the time series have slow linear trends; thus, the SD1 and SD2 descriptors are linear statistics. 

As mentioned by \cite{Arkhypov}, the SD1 and SD2 descriptors are related to the distance from the major and minor axes of the ellipse, respectively 
(for further details, see eqs. 2.5 and 2.6 from \cite{Arkhypov}). SD2 is a weighted combination of low and intermediate frequencies, which portray the long-term characteristics of the signal, and SD1 is strictly a measure of short-term variability. Specifically, using time series data for M dwarf stars, SD1 is related to the magnetic slope or to the photon noise. The contributions of granulation and oscillations are already above the Nyquist limit and clearly below the \textit{Kepler} photon noise. On the other hand, SD2 is related to the dominant period, within which are global trends but not necessarily the trend of the rotation period. 

\subsection{Multifractal indices of magnetic activity}

In the literature, there are several complexity testing techniques available to explore time series \citep{feder1988,mw1969a,mw1969b,mw1969c,Kantelhardt,ihlen}. In general, time series are measured with a wider spectrum of complexity measurements, including nonstationarity, nonlinearity, fractality, stochasticity, periodicity, and chaos \citep{tang}. In this context, linear techniques such as standard deviation do not include the temporal variation at the point-to-point level or multiple lag correlations. For example, the lag-1 Poincar\'e plot does not provide more information about the intermediate time scales that affect the short-lived active regions on characteristic scales of a few hours or days \citep{lanza2003,lanza2004}. A powerful technique to address these assumptions is multifractal analysis \citep{gu2010,tang}\footnote{MATLAB codes for MFDMA analysis can be found on the \texttt{arXiv} version of \cite{gu2010}'s paper: \texttt{https://arxiv.org/pdf/1005.0877v2.pdf}}. As \cite{defreitas2016} proposed, we focus on the multifractal detrending moving average (MFDMA) algorithm. 

\subsubsection{Methodology}
According to \cite{gu2010}, we can summarize the MFDMA algorithm in the following steps:

\begin{itemize}
	\item Step 1: calculating the time series profile:
\end{itemize}
First, we consider a time series $x(t)$ defined over time $t=1,2,3,...,N$, from which construct the sequence of cumulative sums given by
\begin{equation}
\label{eq1}
y(t)=\sum^{t}_{i=1}x(i), \quad t=1,2,3,...,N;
\end{equation}

\begin{itemize}
	\item Step 2: calculating the moving average function of Eq. (\ref{eq1}) in a moving window:
\end{itemize}
\begin{equation}
\label{eq1a}
\tilde{y}(t)=\frac{1}{s}\sum^{\left\lceil (s-1)(1-\theta)\right\rceil}_{k=-\left\lfloor (s-1)\theta\right\rfloor}y(t-k),
\end{equation}
where $s$ is the window size; $\left\lceil (x)\right\rceil$ is the smallest integer not smaller than argument $(x)$; $\left\lfloor (x)\right\rfloor$ is the largest integer not larger than argument $(x)$; $\theta$ is the position index with a range between zero and one; i.e., it describes the delay between the moving average function and the original time series. In the present work, $\theta$ is 0, which refers to a backward moving average; i.e., $\tilde{y}(t)$ is calculated over all past $s-1$ data of the time series.

\begin{itemize}
	\item Step 3: detrending the series by removing the moving average function $\tilde{y}(i)$ and obtaining the residual sequence $\epsilon(i)$ through:
\end{itemize}
	\begin{equation}
\label{eq2}
\epsilon(i)=y(i)-\tilde{y}(i);
\end{equation}

\begin{itemize}
	\item Step 4: calculating the root-mean-square (rms) function $F_{\nu}(s)$ for a segment of size $s$:
\end{itemize}
\begin{equation}
\label{eq3}
F_{\nu}(s)=\left\{\frac{1}{s}\sum^{s}_{i=1}\epsilon^{2}_{\nu}(i)\right\}^{\frac{1}{2}}.
\end{equation}

\begin{itemize}
	\item Step 5: generating the function of $F_{q}(s)$ on the $q$th order:
\end{itemize}
\textbf{\begin{equation}
\label{eq4}
F_{q}(s)=\left\{\frac{1}{N_{s}}\sum^{N_{s}}_{\nu=1}F^{q}_{\nu}(s)\right\}^{\frac{1}{q}}, 
\end{equation}}
for all $q\neq 0$, where the $q$-order is the statistical moment (\textit{e.g.}, for $q$=2, we have the variance); and for $q=0$,

\textbf{\begin{equation}
\label{eq4b}
\ln\left[F_{0}(s)\right]=\frac{1}{N_{s}}\sum^{N_{s}}_{\nu=1}\ln [F_{\nu}(s)], 
\end{equation}}
where the scaling behavior of $F_{q}(s)$ follows a relationship given by $F_{q}(s)\sim s^{h(q)}$, and $h(q)$ denotes the Holder exponent or generalized Hurst exponent.

\begin{itemize}
	\item Step 6: knowing $h(q)$, the multifractal scaling exponent $\tau(q)$ can be computed:
\end{itemize}
\begin{equation}
\label{eq5}
\tau(q)=q h(q)-1.
\end{equation}	

Finally, the singularity strength function $\alpha(q)$ and the multifractal spectrum $f(\alpha)$ are obtained via a Legendre transform, respectively:
\begin{equation}
\label{eq7}
\alpha(q)=\frac{d\tau(q)}{dq}
\end{equation}	
and
\begin{equation}
\label{eq6}
f(\alpha)=q\alpha-\tau(q).
\end{equation}	

For a detailed theoretical description of the MFDMA theoretical background, we strongly recommend the papers by \cite{gu2010} and \cite{defreitas2016}.

\subsubsection{Multifractal indices}

\begin{figure*}
\begin{center}
\subfigure{\includegraphics[width=0.88\textwidth]{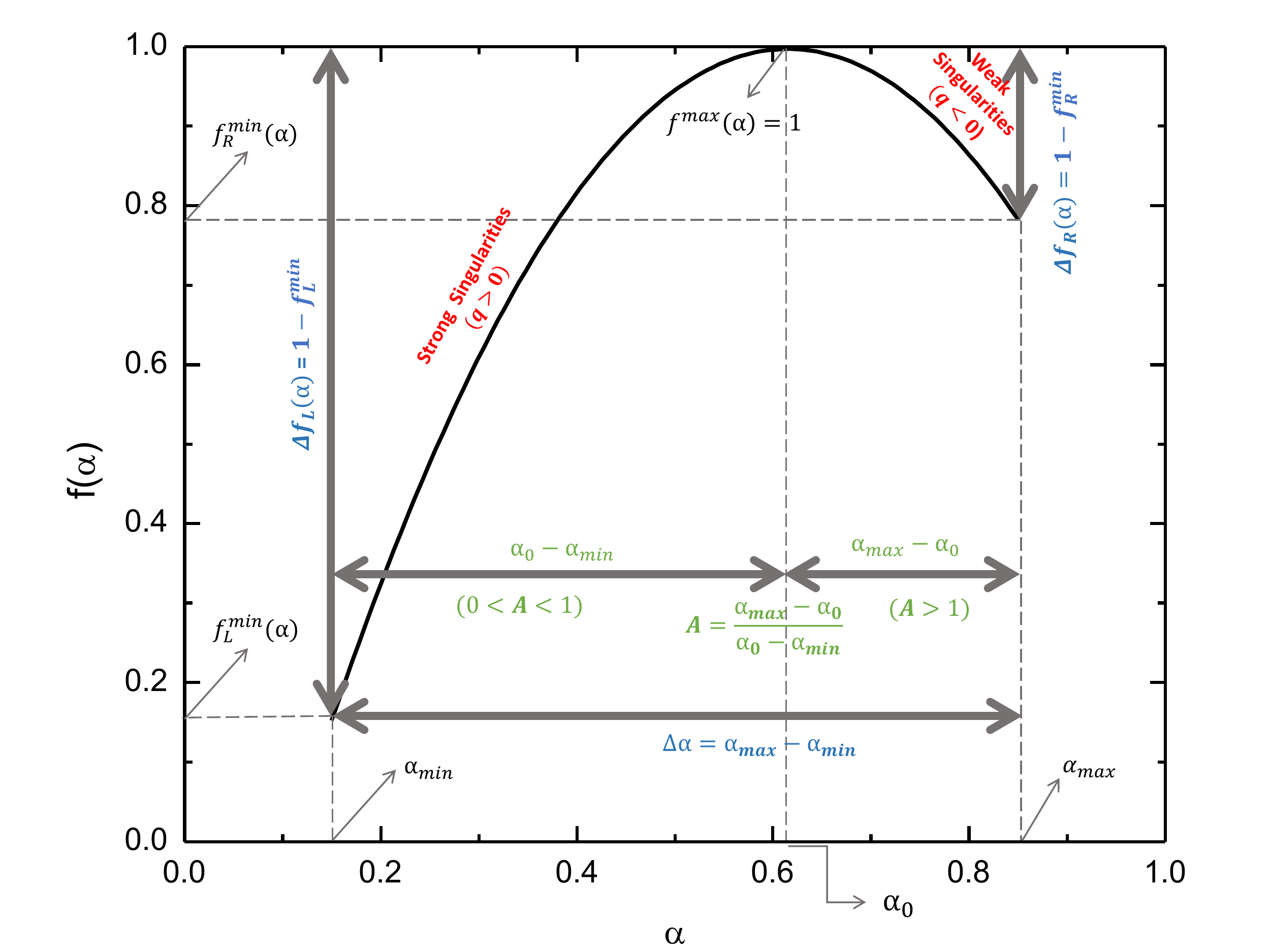}}
\end{center}
\caption{Schematic picture of the multifractal spectrum descriptors with the use of geometric language.}
\label{fig1}
\end{figure*}

Facilitating our discussion of the results, we begin by presenting a typical multifractal spectrum of a time series as the inverse parabolic shape in Fig. \ref{fig1}. For a monofractal time series, $\tau(q)$ is a linear function given by $qH-1$, where $H$ is the global Hurst exponent. For a multifractal signal, $\tau(q)$ is nonlinear, and the multifractal spectrum takes the form presented in Fig. \ref{fig1}. We aim to propose a set of four multifractal descriptors extracted from the spectrum \textbf{$f(\alpha)$} as new stellar magnetic activity proxies, as follows:

\begin{itemize}
	\item \textit{The degree of asymmetry} ($A$), or the skewness in the shape of the $f(\alpha)$ spectrum, may be quantified by the following ratio:
\begin{equation}
\label{eq8}
A=\frac{\alpha_{max}-\alpha_{0}}{\alpha_{0}-\alpha_{min}},
\end{equation}
where $\alpha_{0}$ is the value of $\alpha$ when \textbf{$f(\alpha)$} is maximal. This index presents three shapes according to the value of $A$, which represents asymmetry as right-skewed ($A>1$), left-skewed ($0<A<1$) or symmetric ($A=1$) as illustratedinby Fig. \ref{fig1}. The right endpoint $\alpha_{max}$ and the left endpoint $\alpha_{min}$ denote the extremal values of the singularity exponent, and are associated with the minimum and maximum fluctuation of signal, respectively.
\end{itemize}

\begin{itemize}
	\item \textit{The degree of multifractality} ($\Delta \alpha$) is represented by
\begin{equation}
\label{eq9}
\Delta \alpha=\alpha_{max}(q)-\alpha_{min}(q),
\end{equation}
where $\alpha_{max}$ and $\alpha_{max}$ are the maximum and minimum Holder exponents, respectively, of the statistical distribution of $\alpha$ when $q\rightarrow \mp\infty$ (see Fig. \ref{fig1}). A high value of $\alpha(q)$ indicates that the time series is smooth in that region, and the multifractal strength is consequently lower \citep{defreitas2009}. 
\end{itemize}

\begin{itemize}
	\item  \textit{The singularity ratio} $C$ is characterized by the ratio between $\Delta f_{left}(\alpha)$ and $\Delta f_{right}(\alpha)$ measured in relation to the maximum fractal dimension $f^{max}[\alpha(q=0)]$. The shape of multifractal spectrum, as shown in Fig. \ref{fig1}, represents an asymmetric spectrum. Furthermore, this spectrum can also have either a left or right truncation, as indicated by parameters $\Delta f_{left}(\alpha)$ and $\Delta f_{right}(\alpha)$, and the truncations originate from a leveling of the $q$-order Hurst exponent for negative or positive $q$'s, respectively. In this sense, the index $C$ can be interpreted as a direct measure of truncation, where for $C<1$, the right-hand side is truncated, while for $C>1$, the truncation occurs on the left-hand side. As mentioned by \cite{ihlen}, a long left tail implies that the time series have a multifractal structure that is insensitive to local fluctuations with small magnitudes. On the other hand, a long right tail indicates that the time series have a multifractal structure that is insensivite to local fluctuations with large magnitudes. As shown in Fig. \ref{fig1}, the ratio between the width of the left- and right-sides \textbf{$f(\alpha)$} denotes the degree of the strong and weak singularities as follows:
\begin{equation}
\label{eq10}
C=\frac{\Delta f_{l}(\alpha)}{\Delta f_{r}(\alpha)},
\end{equation}
where the singularity strength $\alpha$ is the inversely proportional to the multifractal spectrum strength \citep{hm}. Moreover, $h$ is a measure of the rate of decay of the fluctuation amplitude; i.e., high values of this exponent denote smoother fluctuations; therefore, the singularity strength is lower. For a time series, the maximum value of \textbf{$f(\alpha)$} is unity. Indeed, this feature reveals that the singularity indicated by \textbf{$\alpha(q)$} is present everywhere in the time series. 
\end{itemize}

\begin{itemize}
\item \textit{The Hurst index} ($H$) can be obtained from the multifractal spectrum through the second-order generalized Hurst exponent $h(q=2)$ (as shown in Fig. \ref{fig1}) \citep[cf.][]{hurst1951,hurst1965}. In particular, $H=h(2)$ for a stationary signal (i.e., with a constant mean and variance expected over time), which is called fractional Gaussian noise (fGn). For the non-stationary case (i.e., with time-dependent variance) with the fractional Brownian motion (fBm), the relationship is $H=h(2)+1$ \citep{movahed,seuront}. As \citep{seuront} proposed, a relevant procedure to distinguish between these two types of processes is to identify the nature of $1/f^{\beta}$ noises, where $\beta$ is a scaling exponent of the Fourier power spectrum. To perform this procedure, we can estimate $\beta$ by determining the slope of a linear trend and identify whether the calculated slope is in an interval of $-1<\beta<1$, where the process is fGn, or whether the interval is $1<\beta<3$, which is characterized as fBm. We can also estimate $\beta$ from the relationship $\beta=2+\tau(2)$, where $\tau(2)$ is the 2nd-order statistical moment mass exponent \citep{ivanov1999}. For example, for $\beta=1$ and 2, we have a flicker noise and random walk, respectively \citep[for further details, see Table 1, p.162, in][]{db2008}. 	
		
As \cite{defreitas2013} reported, when $H$ is between 0.5 and 1, it describes a long-range dependence (LRD) and memory effects on all timescales according to the level of persistence, wherein the time series becomes increasingly periodic as $H$ approaches 1. In contrast, values of $H$ close to zero indicate that the series must change the direction of every sample as white noise. If $H=0.5$, the time series is truly random and uncorrelated data. In particular, the values of $H$ near 0.5 implies a short-range dependence (SRD). A time series with $H<0.5$ can be characterized as anti-persistent; i.e., the signal tends not to continue in the same direction but turns back on itself and gives a less smooth time series \citep{hm}.

\end{itemize}

\begin{deluxetable}{lccccccccccc}
\tabletypesize{\scriptsize}
\tablecaption{Results of the geometric methods. The indices $A$, $\Delta \alpha$, $C$ and $H$ are in the multifractal analysis, whereas $\log SD1[ppt]$ and $\log SD2[ppt]$ were extracted from the Poincar\'e plot. These indices were compared to the rotational period $P_{rot}$ and $\left\langle S_{ph,k}\right\rangle$ in Ref. \cite{mathur2014a}, as shown in the two last columns}.
\tablewidth{0pt}
\tablehead{{\bf Star} & $A$ & $\Delta \alpha$ & $C$ &  $H$ & $\log SD1[ppt]$ & $\log SD2[ppt]$ & $P_{rot}$ & $\left\langle S_{ph,k}\right\rangle$\\
 &  &  &  &  &  &  & (days) &
}
\startdata
{\bf	KIC	2157356}	&	3.025	&	0.699	&	0.233	&	0.444	&	0.06	&	0.91	&	12.9	&	4109.1	\\
{\bf	KIC	2302851}	&	3.066	&	0.768	&	0.303	&	0.404	&	-0.72	&	0.81	&	12.2	&	3934.1	\\
{\bf	KIC	2570846}	&	3.827	&	0.845	&	0.215	&	0.393	&	-0.12	&	1.28	&	10.9	&	11871.9	\\
{\bf	KIC	2574427}	&	2.376	&	0.687	&	0.353	&	0.409	&	-0.3	&	0.37	&	13.4	&	1196.9	\\
{\bf	KIC	2692704}	&	3.228	&	0.670	&	0.259	&	0.475	&	0.02	&	1.16	&	14.8	&	8858.5	\\
{\bf	KIC	2832398}	&	3.946	&	0.781	&	0.181	&	0.463	&	-0.24	&	0.88	&	15	&	4688.7	\\
{\bf	KIC	2834612}	&	3.309	&	0.812	&	0.237	&	0.428	&	-0.11	&	1	&	13.3	&	6391.3	\\
{\bf	KIC	2835393}	&	3.100	&	0.690	&	0.262	&	0.446	&	-0.01	&	0.81	&	15	&	3597.7	\\
{\bf	KIC	3102763}	&	2.977	&	0.714	&	0.272	&	0.454	&	-0.04	&	1.19	&	14.4	&	9310.4	\\
{\bf	KIC	3232393}	&	1.613	&	0.425	&	0.541	&	0.444	&	-0.2	&	0.01	&	14.5	&	404.9	\\
{\bf	KIC	3634308}	&	2.864	&	0.720	&	0.283	&	0.421	&	-0.08	&	0.8	&	12.9	&	3708	\\
{\bf	KIC	3935499}	&	4.272	&	0.842	&	0.216	&	0.257	&	0.08	&	1.45	&	5.2	&	17429.4	\\
{\bf	KIC	4833367}	&	2.541	&	0.571	&	0.302	&	0.438	&	-0.14	&	0.37	&	14.2	&	1135.2	\\
{\bf	KIC	5041192}	&	3.464	&	0.767	&	0.207	&	0.318	&	-0.08	&	0.95	&	10.8	&	5483.9	\\
{\bf	KIC	5096204}	&	3.177	&	0.732	&	0.247	&	0.443	&	-0.62	&	0.32	&	14.8	&	1272.1	\\
{\bf	KIC	5210507}	&	3.374	&	0.867	&	0.253	&	0.333	&	-0.13	&	1.3	&	8.8	&	12345.3	\\
{\bf	KIC	5611092}	&	2.260	&	0.584	&	0.344	&	0.428	&	-0.43	&	0.15	&	14.4	&	711.4	\\
{\bf	KIC	5900600}	&	3.3958	&	0.627	&	0.208	&	0.460	&	-0.26	&	0.64	&	14	&	2194.3	\\
{\bf	KIC	5950024}	&	3.465	&	0.860	&	0.241	&	0.417	&	-0.63	&	0.61	&	14.1	&	2454.5	\\
{\bf	KIC	5954552}	&	4.185	&	0.702	&	0.169	&	0.470	&	-0.44	&	1.02	&	14.9	&	7121.1	\\
{\bf	KIC	5956957}	&	2.813	&	0.680	&	0.264	&	0.443	&	-0.44	&	0.76	&	14.9	&	3825.4	\\
{\bf	KIC	6307686}	&	3.650	&	0.827	&	0.214	&	0.427	&	-0.44	&	0.82	&	13.3	&	3807.7	\\
{\bf	KIC	6464396}	&	3.412	&	0.707	&	0.209	&	0.416	&	-0.21	&	0.93	&	13.2	&	4146.5	\\
{\bf	KIC	6545415}	&	2.831	&	0.865	&	0.266	&	0.324	&	-0.03	&	1.2	&	5.5	&	9018.6	\\
{\bf	KIC	6600771}	&	3.632	&	0.802	&	0.210	&	0.445	&	-0.12	&	1.03	&	13.1	&	6494.6	\\
{\bf	KIC	7091787}	&	2.532	&	0.556	&	0.332	&	0.452	&	-0.08	&	0.39	&	14.1	&	1331.6	\\
{\bf	KIC	7106306}	&	3.513	&	0.900	&	0.233	&	0.416	&	-0.52	&	0.71	&	14.2	&	3194.6	\\
{\bf	KIC	7174385}	&	2.748	&	0.630	&	0.283	&	0.406	&	-0.19	&	0.54	&	14.5	&	2190.2	\\
{\bf	KIC	7190459}	&	2.618	&	0.671	&	0.303	&	0.318	&	-0.27	&	0.64	&	6.8	&	2435.6	\\
{\bf	KIC	7282705}	&	2.486	&	0.661	&	0.337	&	0.428	&	-0.1	&	0.54	&	14.5	&	1842.5	\\
{\bf	KIC	7285617}	&	2.391	&	0.624	&	0.347	&	0.412	&	-0.16	&	0.48	&	13.7	&	1698.3	\\
{\bf	KIC	7534455}	&	3.323	&	0.607	&	0.189	&	0.433	&	-0.32	&	0.56	&	12.1	&	2087.3	\\
{\bf	KIC	7620399}	&	2.299	&	0.555	&	0.343	&	0.440	&	-0.27	&	0.35	&	13.7	&	1170.5	\\
{\bf	KIC	7673428}	&	3.907	&	0.737	&	0.167	&	0.4532	&	-0.12	&	1.07	&	15	&	7213.6	\\
\enddata
\label{tab1}
\end{deluxetable}

\subsection{Correlation methods: Spearman and Pearson coefficients}
To understand the degree of correlation among different parameters (\textit{e.g.}, $H$ and $P_{rot}$), we used two bivariate analysis techniques: (i) Pearson's product moment correlation coefficient and (ii) Spearman's rank correlation coefficient \citep{press}. Qualitatively, the Pearson method measures the strength of the linear relationship between normally distributed variables. However, when the variables are not normally distributed or the relationship between the variables is not linear, the Spearman method is more appropriate \citep{mukaka}.  Quantitatively, Spearman and Pearson coefficients are, respectively, given by 
\begin{equation}
\label{eq11}
r_{S}=1-\frac{6\sum^{n}_{i=1}d^{2}_{i}}{n(n^{2}-1)},
\end{equation}
where $d_{i}$ represents the difference between ranks of variables $x$ and $y$ (for example, $x=A$ and $y=\left\langle S_{ph,k}\right\rangle$) and $n$ is the number of observations,
and
\begin{equation}
\label{eq12}
r_{P}=\frac{S_{xy}}{\sqrt{S_{xx}S_{yy}}},
\end{equation}
where $S$ denotes the covariance. 

 Additionally, if $|r_{S}|>|r_{P}|$, this simply means there is a stronger monotonic than linear relationship. Specifically, a monotonic behavior between the variables can imply a linear relationship and, consequently, the analysis of the results becomes more complex; the presence of outlier may cause this discrepancy between the values of the coefficients. For this case, it is necessary to exclude the outliers, recalculate the coefficients and verify any change. 
 
 In this sense, we want to evaluate whether a correlation exists between $x$ and $y$. To that end, the significance of correlation coefficient can be estimate using a $t$-statistic. First, we specify the null and alternative hypotheses: (i) the null hypothesis $H_{0}:\rho=0$ (there is no association) and (ii) the alternative hypothesis $H_{A}:\rho\neq0$ (a nonzero correlation could exist) for the two-tailed test or $H_{A}:\rho<0$ (a negative correlation could exist) and $H_{A}:\rho>0$ (a positive correlation could exist) as the results for the left- and right-tailed tests, respectively. Second, we calculate the value of the $t$-statistic using the following equation:
\begin{equation}
\label{eq11a}
t_{calculated}=r\sqrt{\frac{n-2}{1-r^{2}}},
\end{equation}
where $n$ represents the sample size and $r$ is the Pearson's correlation coefficient. In addition to the Gaussian distribution, this test is reasonably robust to non-Gaussian data. Third, we use a $t$-table to find the critical value ($t_{critical}$), considering a 95\% confidence level and, consequently, a significance level $\alpha$ equal to 0.05 and 0.025 for each tail related to one and two sided tests, respectively \citep{trauth}. In hypothesis testing, a critical value is a point on the $t$-test distribution that is compared to the calculated $t$-statistic to determine whether the null hypothesis is rejected or not. Finally, we compare the calculated $t$-statistic (see eq.~\ref{eq11a}) to the critical value. In general, if the absolute value of the calculated $t$-statistic is greater than the critical value, then the null hypothesis can be rejected at the 95\% level of confidence in favor of the alternative hypothesis \citep{trauth,press}.

\subsection{Timeout}
In Sections 2.1 and 2.2, we highlighted two words: trend and long-range dependence. We put these special words in bold text to help us understand their meaning in this paper. In addition, we believe a more rigorous definition of this terminology is necessary because of their relationship with the variability in different frequency ranges. A trend is accepted as a part of the time series that changes slowly over time and is broadly defined as a ``long-term change in the mean level'' \citep{db2008}. An alternative to the notion of trends in the time series analysis is to consider the inverse concept of stationarity, such as an fGn-like signal. In fact, a stationary time series is identified as having ``no trend''. However, it is necessary to distinguish between deterministic and stochastic trend-like components. Specifically, a long-range dependence can affect the analysis of this distinction. Several works \citep[\textit{e.g.},][]{db2008,seuront,pg2011} have emphasized that long-range dependence is found in low-frequency variability, which indicates that the autocorrelation function slowly decays and exhibits a scale invariance with the scaling exponent $\beta$. In the astrophysics, LRD can be identified as long-lived features (spots or active regions) and effects because of the differential rotation \citep[cf.][]{daschagas2016}. 

\section{Observations and data preparation}
\begin{figure*}
\begin{center}
\subfigure{\includegraphics[width=0.88\textwidth]{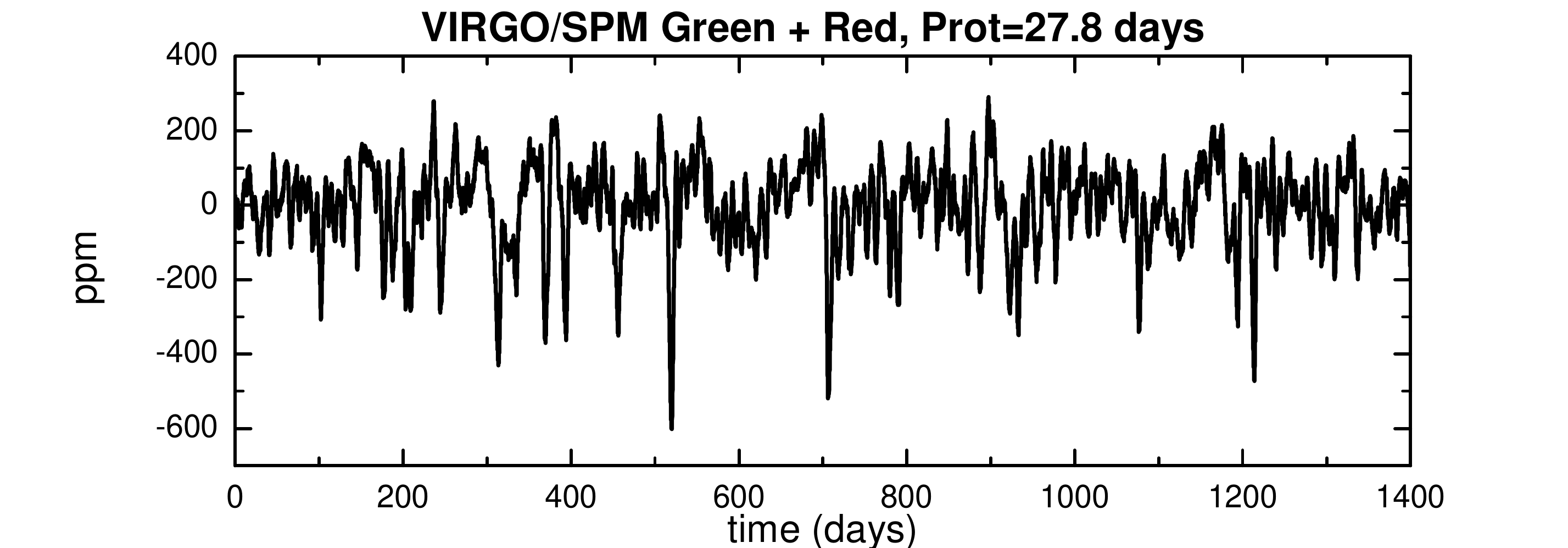}}
\end{center}
\caption{Time series based on the VIRGO/SPM (Green + Red channels) instrument obtained as described in Section 3.}
\label{figSun}
\end{figure*}

In our paper, we use photometric data recorded by the VIRGO/SoHO\footnote{Data are freely available from following website: https://www.ias.u-psud.fr/virgo/} and \textit{Kepler}\footnote{https://www.nasa.gov/kepler/} missions. Our sample is composed of time series of the Sun (see Fig.\ref{figSun}) and a sample of 34 M dwarf stars, which were observed by the \textit{Kepler} mission at a cadence of $\sim$30 min and previously analyzed by \cite{mathur2014a}. These stars are M dwarf stars with $\log T_{eff}$ lower than 3.6, $\log g$ greater than 4.0 and well-defined rotational periods ($P_{rot}<15$ days), as shown in Table 1.
	
The first part of our sample is based in a dataset of Sun continuous observations obtained by Variability of solar IRradiance and Gravity Oscillations (VIRGO) \citep{virgo1,virgo2}. The VIRGO experiment is a component of the payload of the SOHO spacecraft, and is based on four instruments including DIARAD, LOI, PMO6 and SPM. In the present paper, we use the VIRGO data in the green (500 nm) and red (862 nm) bandwidths of the SPM (Sun PhotoMeters) instrument, as proposed by \cite{basri}\footnote{The VIRGO/SPM data can be downloaded from: http://www.spaceinn.eu/data-access/calibrated-sohovirgospm-data/} due to a good approximation with the \textit{Kepler} data. The VIRGO data analyzed in the present work consist of SSI (Spectral Solar Irradiance) time series with a temporal cadence of 1 min and a date range from April 11, 1996 to March 30, 2014, corresponding to solar cycles 23 and 24. To properly compare these results to the stellar case, the time series were averaged into 30-min cadences to match the \textit{Kepler} measurements. This dataset consists of $\sim$18 years of continuous observations; however, as the temporal window of the \textit{Kepler} data is $\sim$4 years, we chose a region in the VIRGO/SPM data with few large gaps from April 22, 1999 to February 20, 2003, within the sun's active phase. The VIRGO/SPM data treatment followed the same procedure adopted by \cite{garcia2005,garcia2011} and \cite{mathur2014a}.

On the other hand, the \textit{Kepler} sample was selected from the calibrated time series processed by the PDC-MAP pipeline \citep{jenkins2010}; a careful treatment was applied to light curves using the so-called co-trending basis vectors provided by the \textit{Kepler} archive \citep{smith,twicken} to remove systematic long-term trends originating from the instruments, detector, or effects caused by re-orientation of the spacecraft every $\sim$90 days. To detect discontinuities and outliers and detrend the data on such short timescales, we applied the method developed by \cite{demedeiros2013}. We also recalculated the rotation periods and found the same results as those reported by \cite{mathur2014a}.

\
\section{Results and Discussion}\label{sec:results}
 As previously mentioned in the introduction, \cite{mathur2014a,mathur2014b} measured the magnetic activity indices using the standard deviation ($S_{ph}$) of the entire time series and the average standard deviation of the subseries defined by $k$-rotation periods (hereafter $\left\langle S_{ph,k}\right\rangle$), where $k=5$ is used. The values of $\left\langle S_{ph,k}\right\rangle$ extracted from \cite{mathur2014a} are reported in Table 1. This procedure divides a time series into macroscopic scales of order or higher than the rotational period. According to the authors, the standard deviation is a good indicator of the global magnetic activity based on photometric modulation and is consequently a classifier of the stellar activity cycle. Specifically, the standard deviation as a function of factor $k$ is only a useful measure to quantify the amount of variation or dispersion in a statistical dataset \citep{feigelson}. However, the long-cadence data used here are filtered, and no signatures of oscillations or granulation are observed above the photon noise. The data are also corrected for photon noise to remove this dependence, taking into account the magnitude of the star as described by \cite{mathur2014a,mathur2014b}. As emphasized by these authors, the $S_{ph}$ index, as calculated, is dominated by the timescales related to rotation and magnetism \citep[for more details:][]{Salabert1, Salabert2}.

\subsection{Results based on the Poincar\'e plot}

\begin{figure*}
\begin{center}
\subfigure{\includegraphics[width=0.48\textwidth]{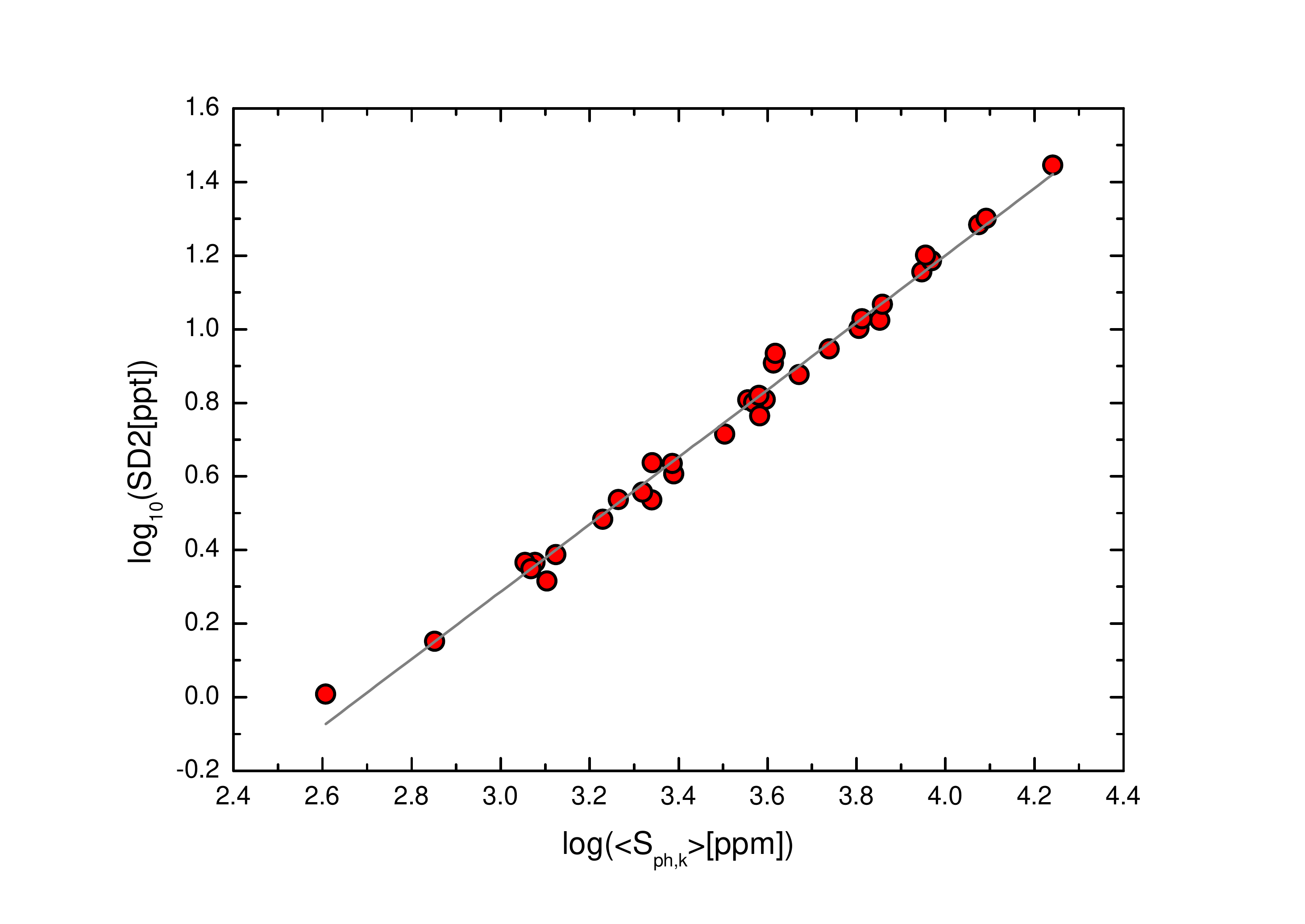}}
\subfigure{\includegraphics[width=0.48\textwidth]{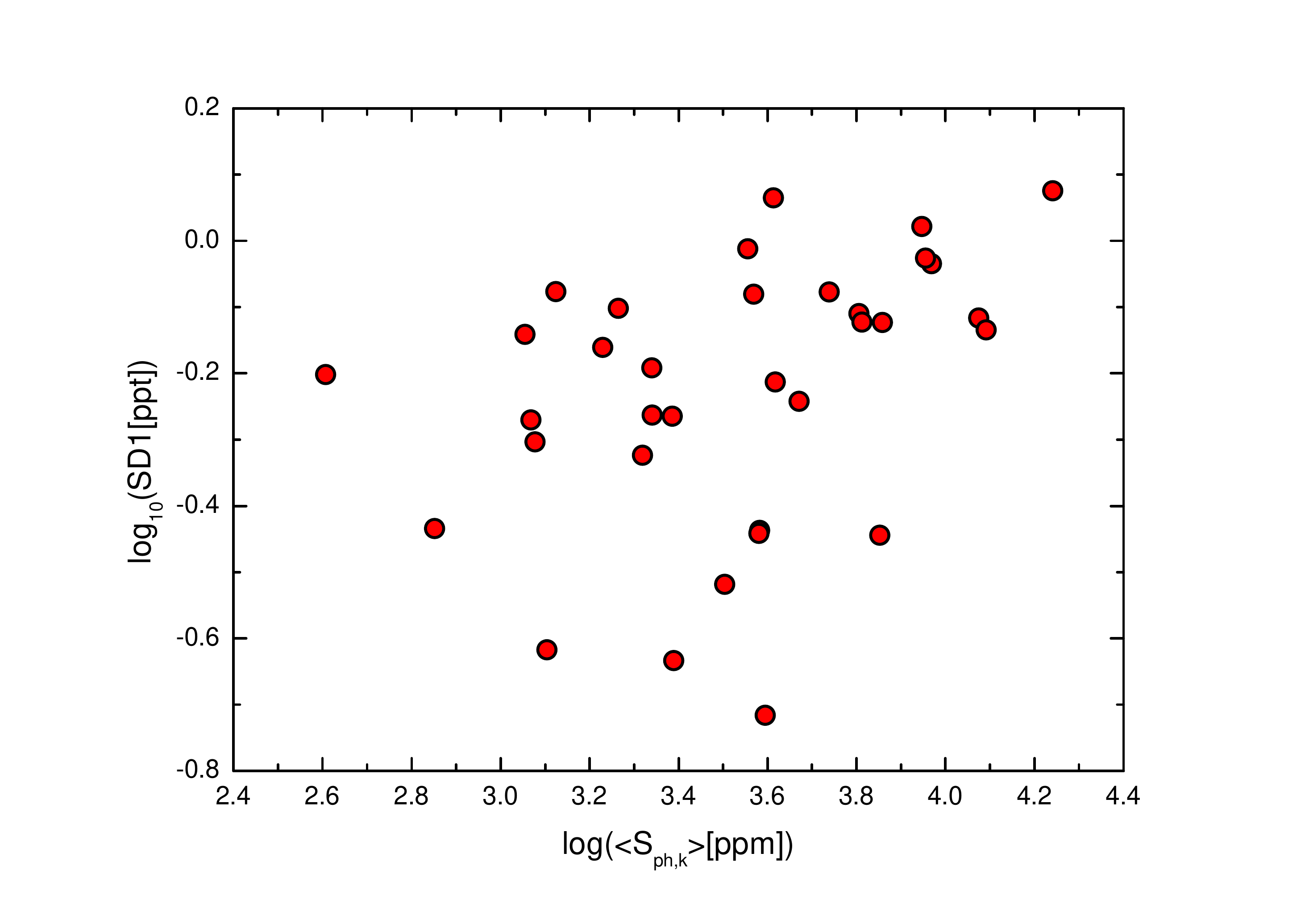}}
\end{center}
\caption{Log-log plots of the $\left\langle S_{ph}\right\rangle$ index versus the SD1 (bottom panel) and SD2 (top panel) descriptors. The solid line corresponds to the linear regression slope.
}
\label{fig0b}
\end{figure*}

As shown in Fig. \ref{fig0b} (left panel), we analyze the sample using the Poincar\'e plot method and find a stronger relationship between the SD2 descriptor and the index $\left\langle S_{ph,k}\right\rangle$ with Spearman and Pearson rank correlations $\sim$1 (the data were tested at a significance level of 5\%). The values of SD2 are summarized in Table 1 (5th column). In contrast, the SD1 descriptor related to short-range variability is weakly correlated with the $\left\langle S_{ph,k}\right\rangle$ index (see left panel from Fig. \ref{fig0b}). The values of SD1 are summarized in Table 1 (6th column). The fact that the $S_{ph}$ index is highly correlated with the SD2 descriptor and not with SD1 provides strong evidence that this index is dominated by long-trend variations and not by short-term variations. With regard to the \textit{Kepler} time series, these short-term variations are the photon noise. All Spearman and Pearson rank correlation coefficient values are presented in Table 2.

\begin{deluxetable}{lccccccccc}
\tabletypesize{\scriptsize}
\tablecaption{Spearman's (1st line) and Pearson's (2nd line) correlation coefficients ($r$).}
\tablewidth{0pt}
\tablehead{ & $A$ & $\Delta \alpha$ & $C$ &  $H$ & $\log SD1$ & $\log SD2$\\
}
\startdata
$\log\left\langle S_{ph}\right\rangle$	&	0.71	&	0.75	&	0.67	&	-0.09	&	0.43 & 0.99	\\
$\log\left\langle S_{ph}\right\rangle$	&	0.76	&	0.76	&	0.72	&	-0.35	&	0.36 & 0.99 \\
\enddata
\label{tab2}
\end{deluxetable}

\subsection{Results based on the multifractal method}
In this paper, we propose a set of multifractal measures as magnetic indices for each star. As a counterpart to the different indices in the literature, our set of indices was defined by the ($A$,\textbf{$\Delta \alpha$},$C$,$H$)-quartet mentioned in Section 2.2 and investigated in \cite{defreitas2013,defreitas2016}. We also apply the multifractal indices proposed in this section to the sample defined by \cite{mathur2014a}. 

\begin{figure*}
\begin{center}
\subfigure{\includegraphics[width=0.48\textwidth]{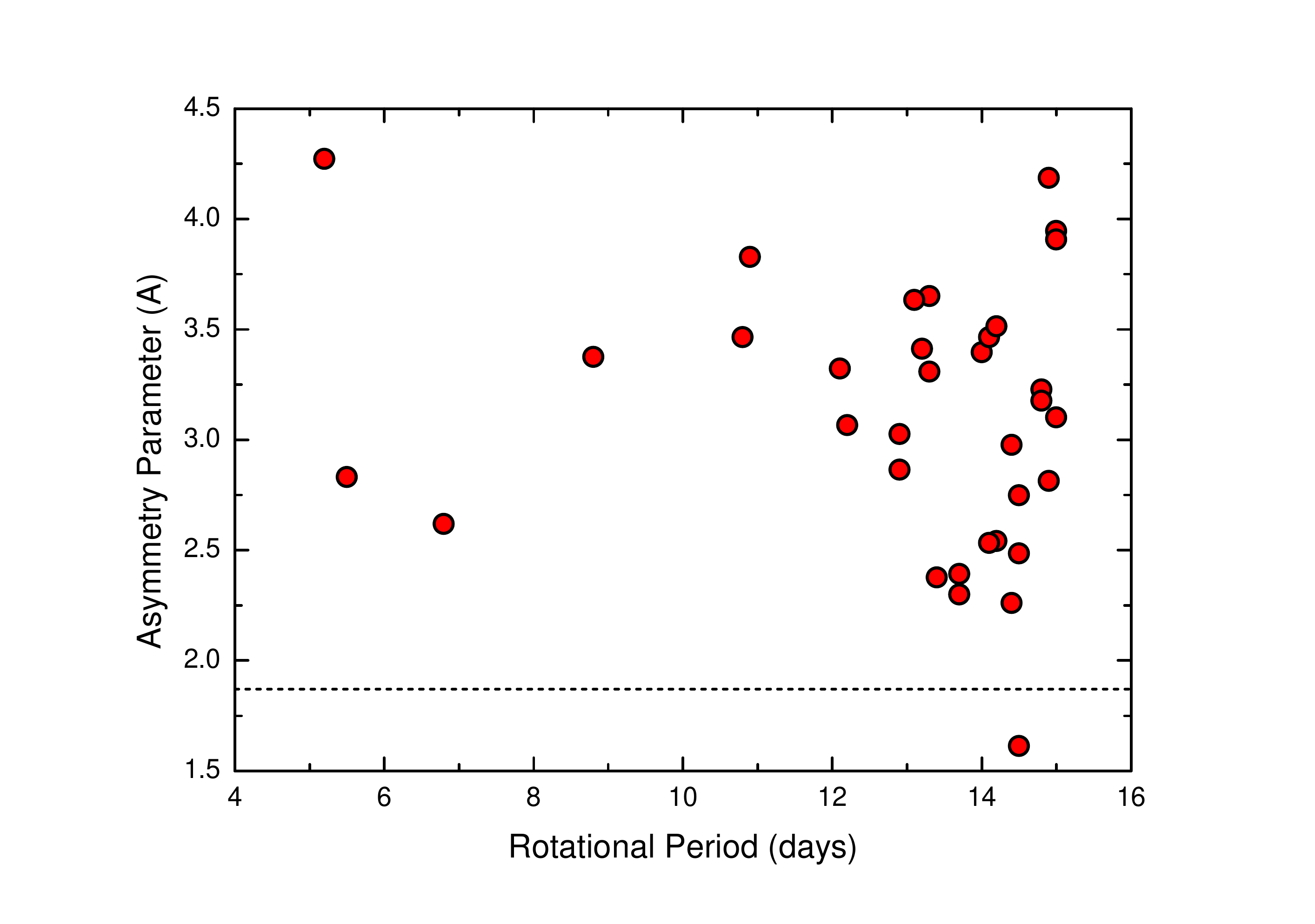}}
\subfigure{\includegraphics[width=0.48\textwidth]{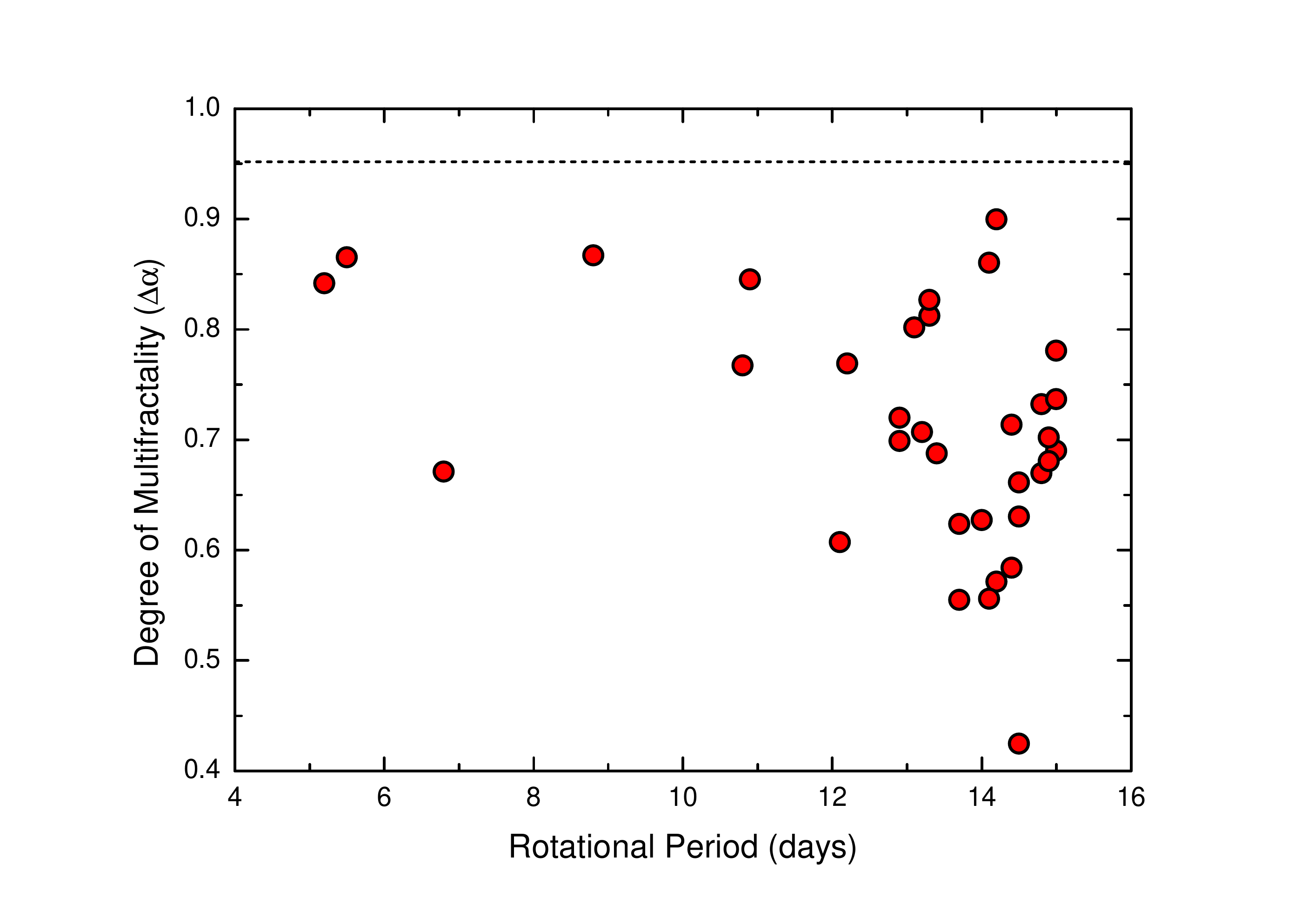}}
\subfigure{\includegraphics[width=0.48\textwidth]{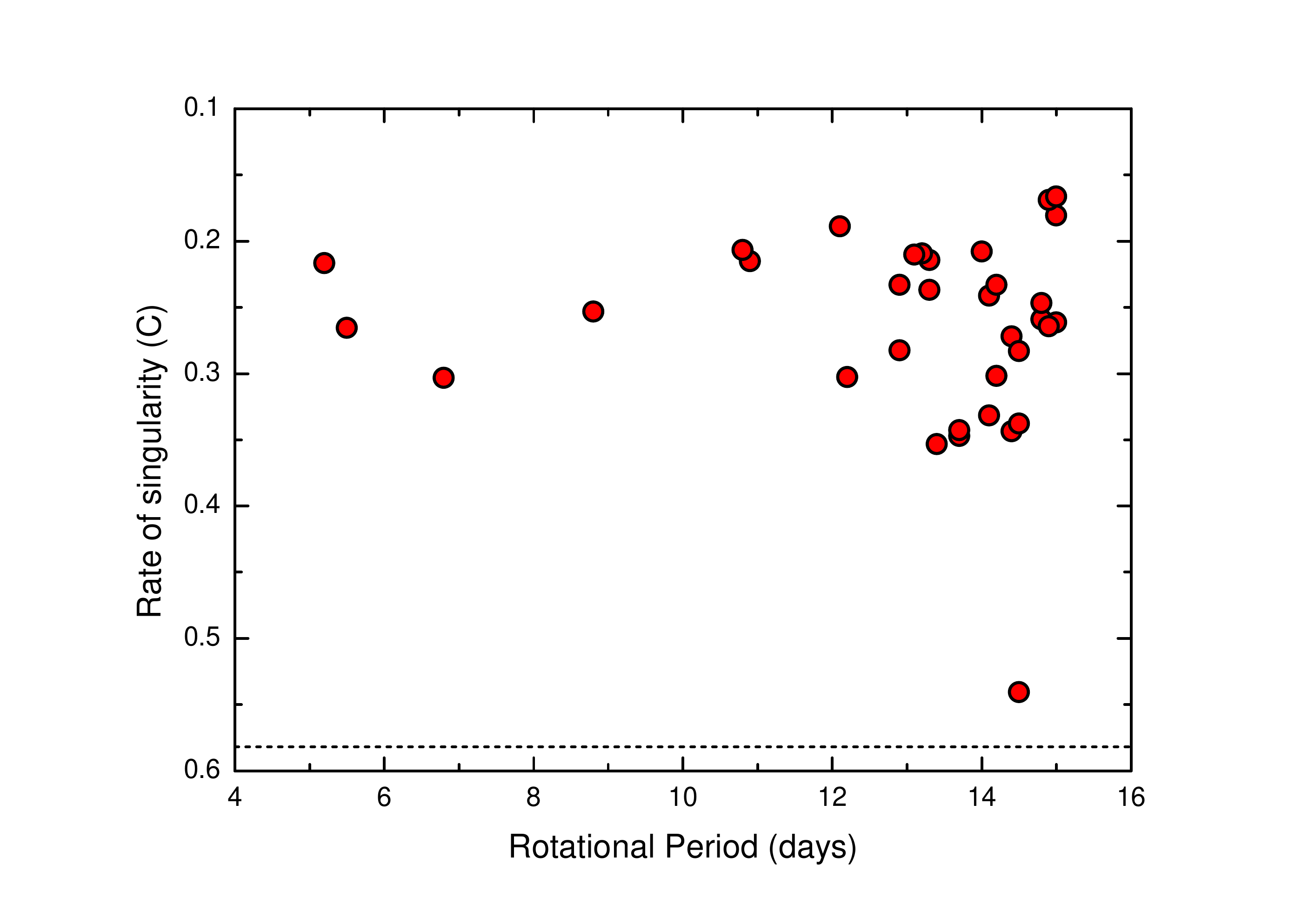}}
\subfigure{\includegraphics[width=0.48\textwidth]{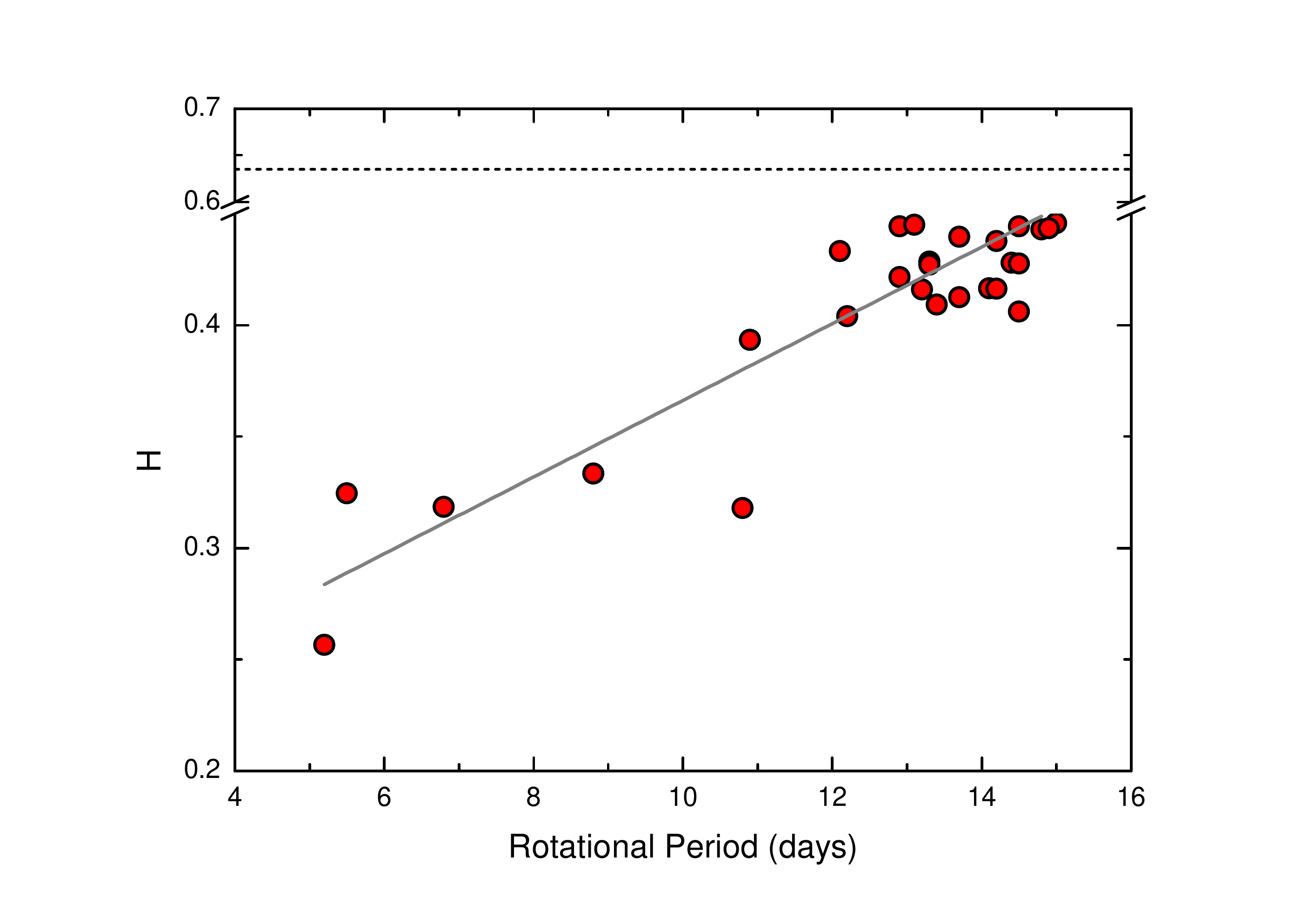}}
\end{center}
\caption{Plotted values of the multifractal indices including the asymmetry parameter ($A$), degree of multifractality ($\Delta \alpha$), rate of singularity ($C$) and global Hurst exponent ($H$), derived from the analysis of the multifractal spectrum as described in Section 2.2 as a function of rotational period. The Sun was analyzed in its active phase. The solar values are represented for the horizontal dashed lines. 
}
\label{fig2}
\end{figure*}

In Fig. \ref{fig2}, we indicate the solar values as horizontal dashed lines. We observe that semi-sinusoidal variations, those illustrated by rotational modulation, do not clearly depend on the indices $A$, \textbf{$\Delta \alpha$} and $C$. In Fig. \ref{fig2} (right bottom panel), we fit the same analytical relationship between the $H$ index and the rotational period proposed by \cite{defreitas2013} (see Eq. 1 from this paper) at the 5\% significance level. As we observe, the global Hurst exponent $H$ grows increases in $P_{rot}$. In addition, this strong correlation between $H$ and $P_{rot}$ supports the results of \cite{defreitas2013}; i.e., the $H$ index is a powerful classifier for semi-sinusoidal time series. Indeed, the first step in the Hurst analysis is to identify whether the dataset is fGn or fBm based on the $\beta$-exponent, as mentioned in Section 2.2.2. For all stars, we found that $\tau(2)\sim -0.02$; hence, $\beta\sim 2$. As a result, all time series can be described as random-walk-like signals with fluctuations evolving mroe slowly than in noise-like time series. Because our values of $h(2)$ are consistently within $1.5\sigma$ of the values of $H$ measured by the rescaled range $R/S$-method in \cite{defreitas2013}, we assume that $H=h(2)$ regardless of the type of signal, although the series is not stationary.

In \cite{mathur2014a,mathur2014b}, the authors suggest that the photometric index $S_{ph}$ is a measure of the magnetic activity. They conclude that a slight anti-correlation between $\left\langle S_{ph}\right\rangle$ and the rotational period can be used to distinguish different levels of magnetic activity and that it shows evidence of long-lived features. We calculate this anti-correlation using the Spearman and Pearson correlation coefficients and find that $r_{S}=-0.27$ and $r_{P}=-0.57$; i.e., $r_{P}\sim 2r_{S}$ is likely due to outliers. 

\begin{deluxetable}{lcccccccccc}
	\tabletypesize{\scriptsize}
	\tablecaption{Matrix of Pearson correlation coefficients showing the simple linear relationship among all the parameters extracted from our sample.}
	\tablewidth{0pt}
	\tablehead{ & $A$ & $\Delta \alpha$ & $H$ &  $\Delta f_{L}(\alpha)$ & $\Delta f_{R}(\alpha)$ & $C$ & $P_{rot}$ & $\log\left\langle S_{ph,k}\right\rangle$ & $\log SD1$ & $\log SD2$
	}
	\startdata
	$\log SD2$	&	0.74	&	0.72	&	-0.36	&	-0.13	&	0.74	&	-0.66	&	-0.46	&	0.74	&	0.41	&	1	\\
	$\log SD1$	&	-0.02	&	-0.09	&	-0.20	&	-0.14	&	-0.12	&	0.00	&	-0.28	&	-0.11	&	1	&	0.41	\\
	$\log\left\langle S_{ph,k}\right\rangle$	&	0.82	&	0.93	&	-0.33	&	-0.01	&	0.99	&	-0.80	&	-0.33	&	1	&	-0.11	&	0.74	\\
	$P_{rot}$	&	-0.17	&	-0.41	&	0.91	&	-0.29	&	-0.35	&	0.10	&	1	&	-0.33	&	-0.28	&	-0.46	\\
	$C$	&	-0.90	&	-0.64	&	0.04	&	0.58	&	-0.76	&	1	&	0.10	&	-0.80	&	0.00	&	-0.66	\\
	$\Delta f_{R}(\alpha)$	&	0.82	&	0.93	&	-0.35	&	0.01	&	1	&	-0.76	&	-0.35	&	0.99	&	-0.12	&	0.74	\\
	$\Delta f_{L}(\alpha)$	&	-0.48	&	0.16	&	-0.38	&	1	&	0.01	&	0.58	&	-0.29	&	-0.01	&	-0.14	&	-0.13	\\
	$H$	&	-0.13	&	-0.42	&	1	&	-0.38	&	-0.35	&	0.04	&	0.91	&	-0.33	&	-0.20	&	-0.36	\\
	$\Delta \alpha$	&	0.71	&	1	&	-0.42	&	0.16	&	0.93	&	-0.64	&	-0.41	&	0.93	&	-0.09	&	0.72	\\
	$A$	&	1	&	0.71	&	-0.13	&	-0.48	&	0.82	&	-0.90	&	-0.17	&	0.82	&	-0.02	&	0.74	\\
	\enddata
	\label{tab2}
\end{deluxetable}

As stated in Section 2.3, the Spearman coefficient indicates that anti-correlations are negligible, whereas the Pearson coefficient indicates that such values are low anti-correlation. Our null hypothesis admits that the correlation between $P_{rot}$ and $S_{ph}$ is zero. For the left-tailed test, the $t$-statistic reveals that, because $\rvert t_{calculated}(-3.888)\rvert>\rvert t_{critical}(-1.692)\rvert$, we can reject the null hypothesis in favor of the alternative hypothesis; i.e., there is a negative correlation between the parameters mentioned. Similarly, for the two-tailed test, the $t$-statistic reveals that because $t_{calculated}=-3.888$ is outside the range of $-2.035<t_{critical}<2.035$, we also can reject the null hypothesis in favor of the alternative hypothesis. As shown in Table 3, the relationships among the different parameters extracted from multifractal analysis, Poincar\'e plot and \cite{mathur2014a}'s analysis were determined through Pearson's correlation coefficients and 10 variables were included included in the correlation matrix, including the relationship between $P_{rot}$ and $\langle S_{ph,k}\rangle$. However, it is not the anti-correlation between $P_{rot}$ and the $S_{ph}$ index that can be used to distinguish different levels of magnetic activity but rather the value of the index itself, as indicated by \cite{mathur2014a,mathur2014b} and \cite{Salabert1,Salabert2}.

\begin{figure*}
\begin{center}
\subfigure{\includegraphics[width=0.48\textwidth]{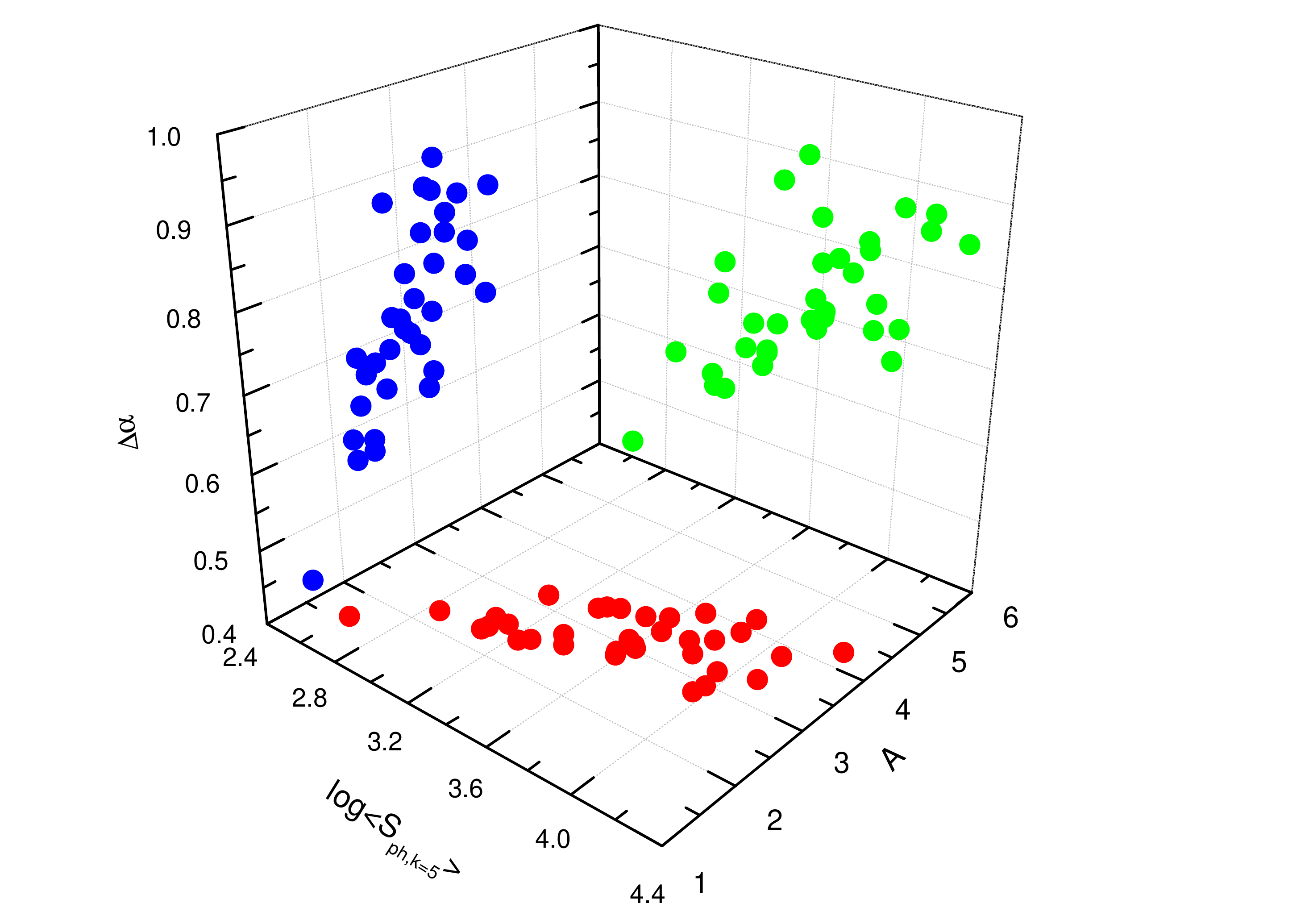}}
\subfigure{\includegraphics[width=0.48\textwidth]{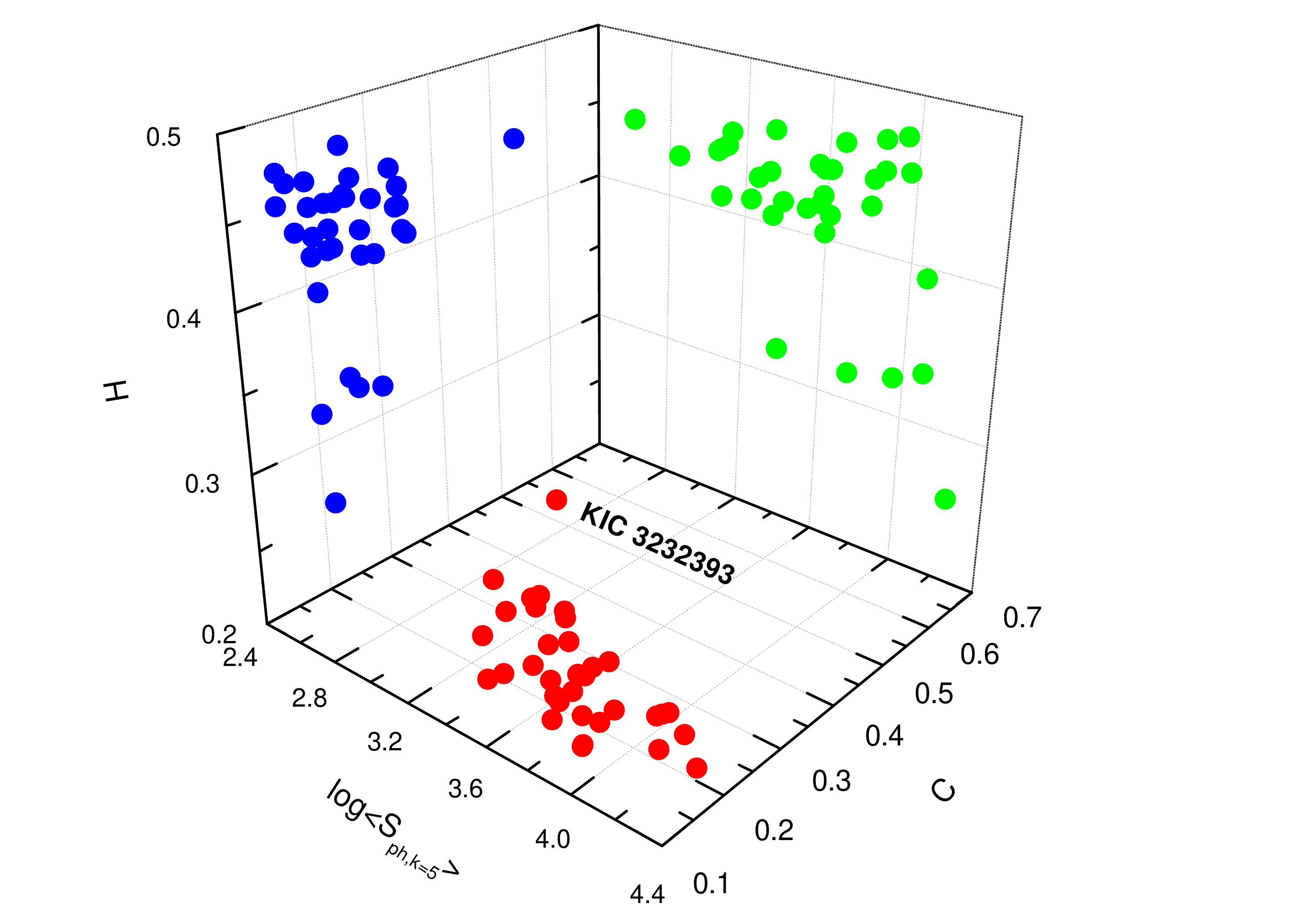}}
\end{center}
\caption{3D plot of projections on the planes of the magnetic indices defined in the space ($A, \log\left\langle S_{ph,k}\right\rangle,\Delta \alpha$) (left panel) and ($C, \log\left\langle S_{ph,k}\right\rangle,H$) (right panel).
The star KIC 3232393 is highlighted.}
\label{fig3}
\end{figure*}

As \cite{mukaka} suggested, the $\left\langle S_{ph}\right\rangle$ index was log-transformed. The reason for this transformation is to build normally distributed observables so that we can use Pearson's coefficient. However, this does not change the results of the previous paragraph. In Figure \ref{fig3}, we defined a point with the vector ($A, \log\left\langle S_{ph,k}\right\rangle,\Delta \alpha$) (left panel) and ($C, \log\left\langle S_{ph,k}\right\rangle,H$) (right panel) from the projections on the planes. Thus, we constructed a 3D plot in which point represents a time series with a particular rotational period, as indicated in Fig. \ref{fig2}. According to the projections ($A, \log\left\langle S_{ph,k}\right\rangle$) and ($\log\left\langle S_{ph,k}\right\rangle,\Delta \alpha$) of the left panel from Fig. \ref{fig3}, the $\left\langle S_{ph}\right\rangle$ index is strongly linked to the geometric properties of the multifractal spectrum \textbf{$f(\alpha)$} by parameters $A$, $\Delta \alpha$ and $C$. 

In Table 2, we verify that the Spearman and Pearson coefficients between $\left\langle S_{ph,k}\right\rangle$ and the $A$ and $\Delta \alpha$ indices are significantly higher than those for the index $C$. This disparity occurs because there is an outlier, identified as star KIC 3232393, which presents the highest singularity ratio $C$ but does not affect the other correlations. After excluding this star, we recalculate the coefficients and obtain\textbf{ $r_{S}=-0.62$} and \textbf{$r_{P}=-0.64$}. 

According to \cite{noyes}, there is a correlation between the observed flux indices (Ca II HK flux and $R^{'}_{HK}$) and the rotation period. We verified that the multifractal $H$ index has the strongest correlation with $P_{rot}$. As proposed by \cite{defreitas2013}, the strong correlation between the $H$ index and the rotation period would define $H$ as a measure of the intrinsic memory in the light curve affected by semi-sinusoidal variations. On the other hand, the panels of Figure \ref{fig2} are in agreement with the results of \cite{mathur2014a}, in which faster rotators are more active. Qualitatively, in the left panel of Fig. \ref{fig3} and in the ($\log\left\langle S_{ph}\right\rangle$-C)-plane from the right panel of same figure, the correlation between $\left\langle S_{ph}\right\rangle$ and the ($A$, $\Delta \alpha, C$)-triplet reveals that our multifractal indices are related to the magnetic activity of the M dwarfs analyzed here.

Table 1 shows that all $H$ values are below $0.5$, which indicates that the fluctuations in the amplitude of the photometric flux are anti-correlated. In other words, an amplitude has decreased in the past is more likely to increase than decrease in the future. According to the Central Limit Theorem, for long time series, we expect values of $H\sim 0.5$ because the memory between two points decreases with an increasing number of data points. These results can give us a clearer idea about the physical implications of the photometric index that \cite{mathur2014a} used; consequently, we suggest that a multifractal framework can give us a new approach to \cite{mathur2014a}'s index.

\subsection{The $\left\langle S_{ph,k}\right\rangle$ index and the source of its geometric dependence}

Because the $\left\langle S_{ph,k}\right\rangle$ index is strongly correlated with the geometric indices $A$, $\Delta \alpha$ and $C$, we decided to investigate the changes in the curve of the \textbf{$f(\alpha)$} spectrum by comparing the indices calculated from the original series with those obtained from the shuffled and surrogate series. This procedure finds the possible source(s) that affect the profiles of $A$, $\Delta \alpha$ and $C$ and consequently suggest the abovementioned strong correlation. In general, the shuffled series methods removes any temporal correlations that eliminate the memory of the system, but the method does not affect the probability distribution function (PDF). The surrogate procedure eliminates non-linearities and preserves only the linear properties of the actual series \citep{Norouzzadeha,defreitas2016}.

\begin{figure*}
\begin{center}
\subfigure{\includegraphics[width=0.88\textwidth]{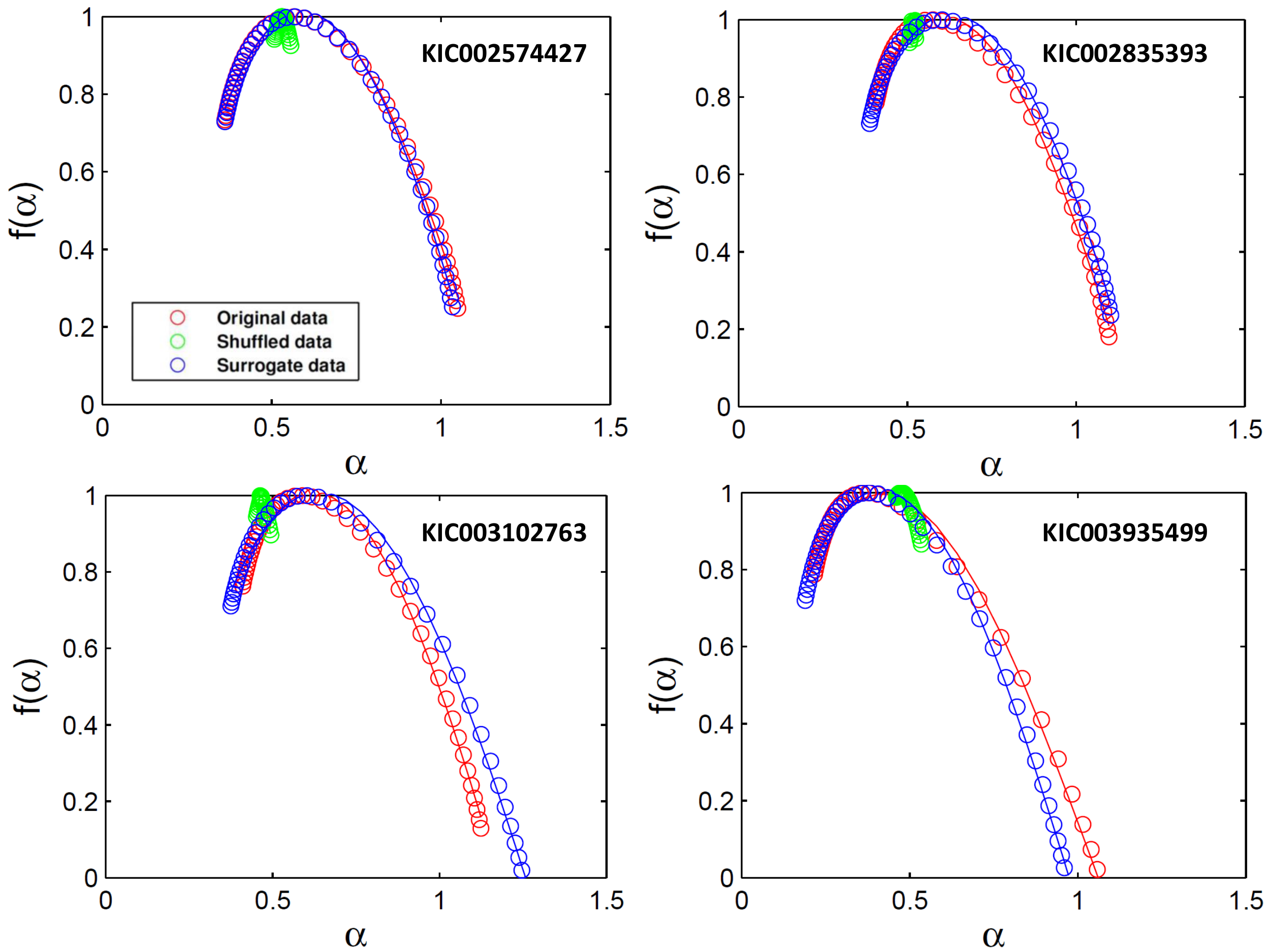}}
\end{center}
\caption{Multifractal spectra of \textbf{$f(\alpha)$} versus $h$ of the original (red), shuffled (green) and surrogate (blue) time series. The spectra were created by the MFDMA algorithm. 
}
\label{fig4}
\end{figure*}

In all multifractal spectra, the ($A$, $\Delta \alpha$, $C$)-triplets in the original time series are higher than their shuffled partners (see the red circles in Fig.\ref{fig4}). Only a portion of the spectra from our sample are represented in Fig.\ref{fig4}. In contrast, there is no difference between the indices from the original and surrogate time series, which indicates that the structural properties of the signal are essentially linear; i.e., ($A$, $\Delta \alpha$, $C$)$_{surrogate}\sim$($A$, $\Delta h$, $C$)$_{original}$. In this context, we observe a meaningful effect of random shuffling in the spectra \textbf{$f(\alpha)$}. We observe a weak multifractal effect in the shuffled series from the generated \textbf{$f(\alpha)$} spectra; i.e., the triplet ($A$, $\Delta \alpha$, $C$)$_{shuffled}<$($A$, $\Delta \alpha$, $C$)$_{original}$. This behavior probably arises from the heavy-tailed distribution of the time series data.

In conclusion, the geometric dependence of the $\left\langle S_{ph,k}\right\rangle$ index originates from the multifractality of both the correlations and the PDF. This result reinforces the strong correlation between $\left\langle S_{ph,k}\right\rangle$ and SD2 descriptor.

\subsection{Possible effects derived of the inclination angle}

We have not taken into account the possible effects due to the inclination angle of the rotation axis in relation to the line of sight. As investigated by \cite{vazquez} and reported by \cite{mathur2014b}, the angle of inclination of the star, which can be estimated using a combination of the stellar radius ($R$), projected equatorial velocity ($\sin i$) and rotation period, is \textcolor[rgb]{1,0,0}{a} relevant parameter due to effects it can have on observations of magnetic cycles. Unfortunately, we can only derive the $\sin i$ value for KIC 6464396. For this star, $i\sim 90^{0}$ \citep{prsa}.

We can estimate the median rotation velocity values from $R_{KIC}$, the \textcolor[rgb]{1,0,0}{\textit{Kepler}} stellar radius, and $P_{rot}$ through the relationship $\bar{v}=2\pi\left\langle R_{KIC}/P_{rot} \right\rangle $. In addition, as the triplet ($A$, $\Delta \alpha$, $C$) is linked to $\left\langle S_{ph,k}\right\rangle$, we can assume that the triplet also represents a lower limit of the photospheric activity \citep{Salabert2}. In this sense, spectroscopic measurements are necessary for estimating the effect of $i$ over our sample to reach a conclusive result.

\subsection{Age-rotation-activity relationship and the astrophysical meaning of the $H$ index}

\begin{figure*}
\begin{center}
\subfigure{\includegraphics[width=0.88\textwidth]{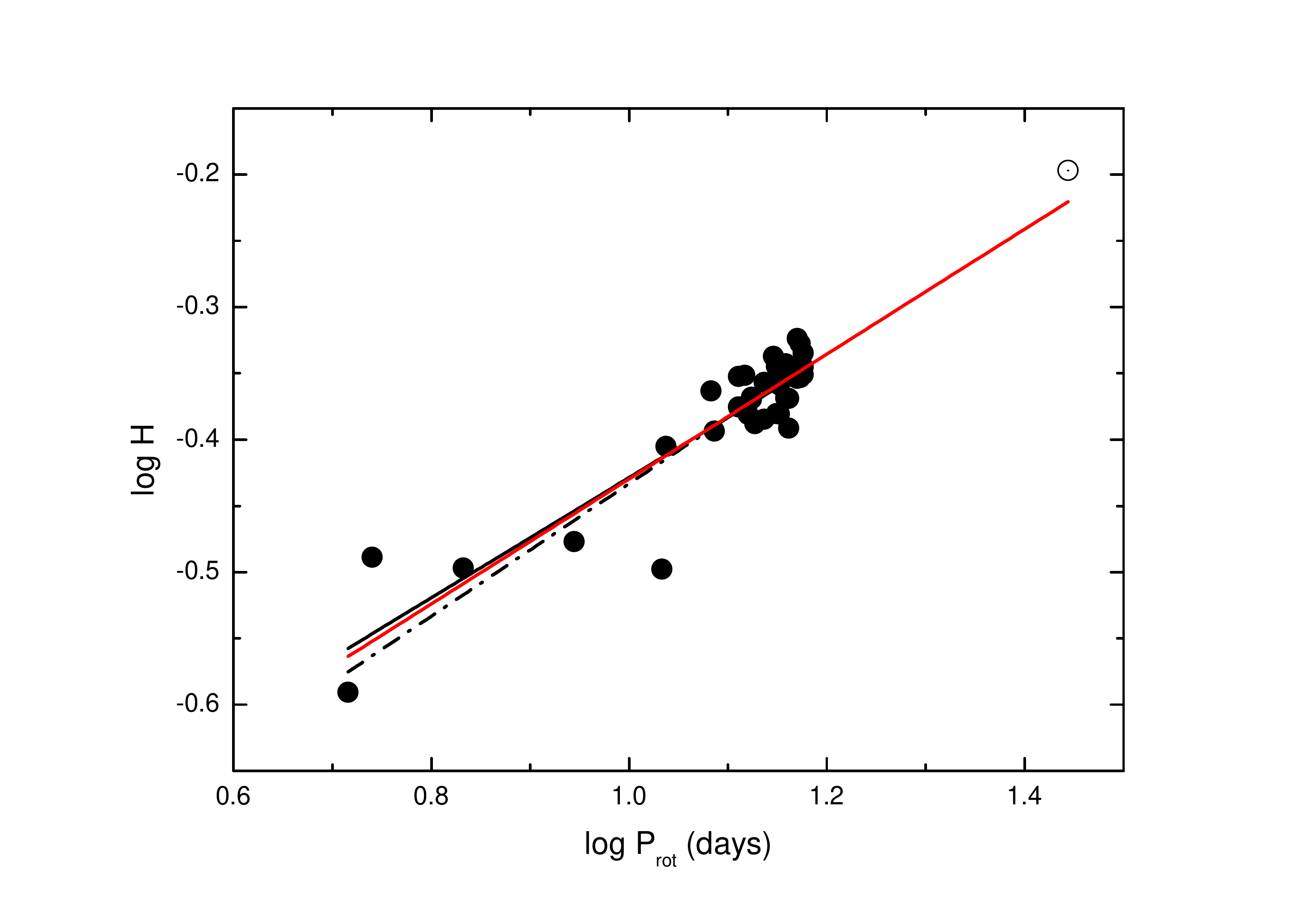}}
\end{center}
\caption{The Hurst exponent measured via multifractal procedure as a function of rotation period.
The black solid line gives us the following log-log relationship: $\log H=(0.45\pm 0.04)\log P_{rot}-(0.88\pm 0.04)
$. The dotted line shows the log-log relationship from \cite{sku}. Considering all the samples (34 stars $+$ the Sun), the red solid line gives us the following log-log relationship: $\log H=(0.47\pm 0.03)\log P_{rot}-(0.90\pm 0.03)
$. The Sun is represented by the symbol $\odot$. For both the reduced $\chi^{2}$, value of the fit is
$\sim 10^{-3}$. 
}
\label{fig5}
\end{figure*}

Since the pioneering study by \cite{sku}, the age-rotation-activity connection has been used to investigate the evolution of the stellar angular momentum and its implications for magnetic activity levels driven by the stellar dynamo \citep{kawaler,defreitasmnras}. In the present context, the rotation-$H$ relationship suggests a clear similarity with a typical relationship known as the age-rotation relationship. Instead, this type of connection is expected in models that involve a link between rotation and a magnetic activity index proposed by \citep{karoffb,metcalfe}. As shown in Fig. \ref{fig5} and inspired by the age-rotation relationship, we suggest that the rotation-$H$ relationship is better described as a power-like law of $H=aP_{rot}^{b}$, where $a$ is a normalization constant and $b$ is the scaling exponent. In Fig. \ref{fig5}, the log-$H$--log-$P_{rot}$ relationship is represented by the dark solid line given by
\begin{equation}
\label{eq14}
\log H=(0.45\pm 0.04)\log P_{rot}-(0.88\pm 0.04).
\end{equation}

The important parameter to compare here is the exponent $0.45\pm 0.04$. This slope is consistent with the original result of 0.5 in \cite{sku}. Including the Sun in the plot, we verify that the relationship is maintained:
\begin{equation}
\label{eq15}
\log H=(0.47\pm 0.03)\log P_{rot}-(0.90\pm 0.03).
\end{equation}

It is worth noting that the reason for this relationship is the different spin-down timescales for stars of different masses, which contradicts the results based on the $\left\langle S_{ph,k}\right\rangle$ index \citep[see Fig. 3 in][]{mathur2014a}. Compared to the multifractal theoretical background, this result shows that the evolution of the angular momentum is a function of the age or mass as well as the dynamics of magnetic activity over different timescales, which is characterized by the $H$ index here.
\\

\section{Concluding remarks}
In the previous section, after proving the correlation between the $H$ index and the rotation period, we verified the possible correlations between the ($A$,$\Delta \alpha$,$C$,$H$)-quartet and the index defined by \cite{mathur2014a}. This analysis suggests that the multifractal properties of this quartet are highly correlated with the $\left\langle S_{ph}\right\rangle$ index, though the $H$ index is strongly linked to the rotational period. Moreover, the ($A$,$\Delta \alpha$,$C$)-triplet is strongly correlated with this index. Because of this interrelationship, we conclude with full confidence that the $\left\langle S_{ph}\right\rangle$ index is a measure of the effects of both temporal correlation and a heavy-tailed PDF, both of which affect the geometry of the multifractal spectrum.

We have compared the measures of both the $\left\langle S_{ph}\right\rangle$ and $H$ indices. Our result shows a strong relationship between the stellar rotation period from the analysis of the light curve and the values of the $H$ index, as suggested by \cite{defreitas2013}. It is worth noting that this result strengthens the general proposition that the $H$ index is related to the evolution of stellar angular momentum and magnetic activity. Because of the properties of $H$, the stellar rotation is associated with the degree of persistence of the signal.

From a physical viewpoint, perhaps because of this universal feature, we understand that multifractality does not imply a consequence of particular physical properties but rather represents a more general behavior of complex systems that have similar entities. In conclusion, the universality of the multifractal nature provides us with a more robust and deeper statistical ensemble than the traditional and more conservative approaches that dominate the usual methods in the astrophysical literature, which characterize stellar photometric variability in a notably limited manner.

\acknowledgments
DBdeF acknowledges his wife (Nara) and sons (Guilherme and Helena) for their warm and
continuous support beyond the realm of science. He also acknowledges financial support 
from the Brazilian agency CNPq-PQ2 (grant No. 306007/2015-0). ICL acknowledges a postdoctoral fellowship from the Brazilian agency CNPq (Science Without Borders program, Grant No. 207393/2014-1). This paper includes data collected by the \textit{Kepler} mission. Funding for the \textit{Kepler} mission is provided by the NASA Science Mission directorate.  All data presented in this paper
were obtained from the Mikulski Archive for Space Telescopes
(MAST). We also gratefully acknowledge the SoHO team for providing the data used in this work. The VIRGO instrument on board SoHO is a cooperative effort of scientists, engineers, and technicians, to whom we are indebted. We thank the anonymous referees for suggestions that helped improve our
work. Research activities of the Astronomy Observational and Astrostatistics Board of the Federal University of Rio Grande do Norte and Federal University of Cear\'a are supported by continuous grants from the Brazilian agency CNPq. We also acknowledge financial support from INCT INEspa\c{c}o/CNPq/MCT. Research activities of the Astronomy Observational and Astrostatistics.

\end{document}